\documentclass{pasj00}

\begin{document}
\SetRunningHead{K. Nakazawa et al.}{Hard X-Rays from Groups of Galaxies}
\Received{2006/09/25}
\Accepted{2006/11/21}
\title{Hard X-Ray Properties of Groups of Galaxies as Observed with ASCA}
\author{Kazuhiro \textsc{Nakazawa}}
\affil{Institute of Space and Astronautical Science, JAXA,	3-1-1 Yoshino-dai, Sagamihara, 
Kanagawa 229-8510}
\email{nakazawa@astro.isas.jaxa.jp}
\author{Kazuo \textsc{Makishima}
 \thanks{Also with RIKEN, Wakou-shi, Saitama 351-0198, Japan}}
\affil{Department of Physics, University of Tokyo, 
7-3-1 Hongo, Bunkyo-ku, Tokyo 113-0033}
\and
\author{Yasushi {\sc Fukazawa}}
\affil{Department of Physical Science, Hiroshima University, 1-3-1 Kagamiyama, Higashi-hiroshima, 
Hiroshima 739-8526}

\maketitle

\begin{abstract}
X-ray spectra of groups of galaxies,
obtained with the GIS instrument onboard ASCA, 
were investigated for diffuse hard X-rays in excess of 
the soft thermal emission from their inter-galactic medium (IGM).
In total, 18 objects with the IGM temperature
of 0.7--1.7 keV were studied,
including HCG 62 in particular.
Non X-ray backgrounds in the GIS spectra
were carefully estimated and subtracted.
The IGM emission was represented by up to two temperature thermal models,
which was determined in a soft energy band
below 2.5 keV mainly by the SIS data. 
When extrapolated to a higher energy range of 4--8 keV, 
this thermal model under-predicted the
background-subtracted GIS counts in HCG 62 and RGH 80 
by $> 2\sigma$ significance, even though the background uncertainties
and the IGM modeling errors are carefully accounted.
A hard excess could be also present in NGC 1399.
The excess was successfully explained
by a power-law model with a photon index $\sim 2$,
or a thermal emission with a temperature exceeding $\sim3$ keV.
In HCG 62, the 2--10 keV luminosity of the excess hard component 
was found to be $5.5\times10^{41}$ erg s$^{-1}$ at 2--10 keV, 
which is $\sim 30$ percent of the thermal IGM luminosity in 0.7--2.5 keV.
Non-thermal and thermal interpretations of this excess components are discussed.
\end{abstract}

\section{Introduction}
\label{chap:intro}

Clusters  of galaxies shows rich evidence for
huge energy input within their vast inter-galactic space.
Mega-parsec scale radio halos observed in many rich clusters 
provide direct evidence of GeV electrons accelerated within them
(e.g. Feretti, Giovannini 1996).
Significant temperature variations in the cluster 
hot gas, detected by ASCA (e.g. Furusho et al.~2001)
as well as Chandra and XMM-Newton 
(e.g. Markevitch et al.~2003 ; Briel et al.~2004),
can be interpreted as relics of energy dissipation associated with cluster mergers.
Another striking feature is so called ``cavity" 
in the cluster centers (e.g. Birzan et al.~2004),
which is suggestive of an energy input at a level as high as $10^{58-60}$ erg.
These new violent features of clusters of galaxies challenge
the classical view of a single phase hot plasma hydorstatically
filling the gravitational potential formed by a dark matter halo.
The intra-cluster volume, for example,  may harbor 
plasma components much hotter than the virial temperature
and/or a significant amount of non-thermal particle population.

The same stroy may apply also to groups of galaxies, 
the subject of the present paper.
They hosts a fair amount of hot gas,
called inter-galactic medium (IGM) with a temperature of about 1 keV
(see e.g. Mulchaey et al.~1996).
X-ray emission from the IGM is dominated by 
Fe-L shell lines which appear in the soft X-ray range around 0.6-1.4 keV.
Since the emissivity of IGM is very low in energies above 2 keV, 
we can search this ``hard" energy region of the X-ray spectra of groups of galaxies
for any additional harder emission component, such as thermal signals from hotter plasmas
mixed in the IGM or non-thermal emission from accelerated particles.

\begin{figure}[htbp]
	\begin{center}
	\FigureFile(80mm,50mm){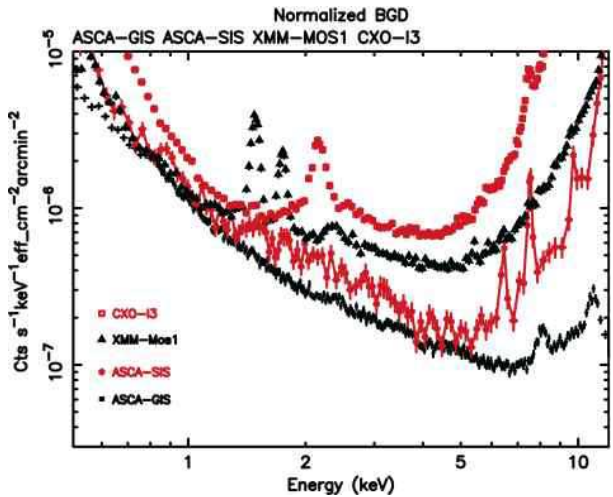}
	\end{center}
\caption{Background spectra of imaging X-ray detectors per sky area,
normalized also to the mirror effective area.
Results of the ASCA GIS and the SIS (this work) are compared with those of
the Chandra ACIS I 3 and the XMM-Newton EPIC mos 1 (from template background files).
Spectra are extracted from a central $15'$ circular region for the GIS,
one full chip for the SIS and ACIS I 3, and a central $8'.5$ circular region for EPIC mos.
The effective area is for a point source on the nominal aim-point of each detector.
The cosmic X-ray background is included.
}
\label{fig:intro:GISmerit}
\end{figure}

The ASCA mission (Tanaka et al.~1994), operated from 1993 to 2000, 
was equipped with four X-ray mirror optics
covering the energy range up to $\sim 10$ keV.
Although their angular resolution was limited,
the total effective area of ASCA at 6 keV is larger than that of Chandra.
Furthermore, the GIS experiment
(Makishima et al.~1996; Ohashi et al.~1996) onboard ASCA 
is characterized by its very low and stable background,
together with a wide field of view ($\sim 45'$ in diameter)
and a high quantum efficiency toward $\sim 10$ keV.
Thanks to these properties, the GIS background level
normalized to the effective area and sky area, shown in figure \ref{fig:intro:GISmerit},
has been the lowest among the X-ray detectors
with imaging spectroscopic capabilities up to 10 keV.
Therefore, data from the GIS is least affected by background uncertainties
(both statistical and systematic), which strongly limit the sensitivity to 
largely ($\gtrsim 5'$) extended emission in the hard X-ray band above $\sim 4$ keV.
The power of the GIS detector is demonstrated, for example, by the detection of 
non-thermal inverse Compton emission from the radio lobe of Fonax-A (Kaneda et al.~1995).
As a result, the ASCA GIS archive is expected, even today, to provide the best opportunity
to search for excess hard X-ray signals from groups of galaxies.
Its main drawback, namely the poor sensitivity for exclusion of contaminating 
point sources, can be compensated for by referring to the public data
with higher angular resolution, such as ROSAT and Chandra.

Based on this idea, Fukazawa et al.~(2001) analyzed the ASCA data of 
a compact galaxy group HCG 62,
and discovered evidence for a hard X-ray excess above the thermal IGM emission.
Their approach is considered applicable 
not only to this particular group, but also to almost all near-by low temperature groups.
In this paper,
we hence study 18 near-by low temperature galaxy groups in the ASCA archive data.
In section 2, we present the list of targets with their selection criteria.
Section 3 describes the analysis method we employed.
Results from the analysis is summarized in section 4.
Discussion is presented in section 5, followed by a brief summary in the last section.
Throughout the present paper, we assume the Hubble constant
to be $H_0 = 75$ km s$^{-1}$ Mpc$^{-1}$.
All the errors refer to 90 \% confidence levels, unless otherwise noted.

\section{Target selection}
\label{chap:target}

We surveyed all the ASCA archival data for groups of galaxies 
that are appropriate for our purpose.
To separate any additional hard components from the IGM emission 
within the ASCA band-pass, we need targets with soft and bright IGM emission.
In addition, targets having near-by hard X-ray sources must be avoided, 
because the wide wing of the point spread function (PSF) of the ASCA mirror
degrades the sensitivity. We must avoid even hard sources outside the GIS 
field of view, because they produce stray lights.

We selected objects which has an IGM temperature lower than 1.7 keV,
and the 0.5--10 keV flux higher than $\sim 1 \times 10^{-12}$ erg s$^{-1}$ cm$^{-2}$.
We excluded objects with a luminosity less than
$5 \times 10^{41}$ erg s$^{-1}$,
most of which are isolated elliptical galaxies rather than groups: 
this is to reduce the hard X-ray 
contribution from unresolved discrete sources, such as low mass X-ray binaries (LMXBs).
We also examined the ROSAT all sky map and
other literature for the absence of nearby ($<2~^{\circ}$) active galacitc nuclei (AGN).

The sample thus selected consists of 18 groups, including HCG 62.
We list them in table \ref{tbl:obs:tarlist} together with their optical properties.
Although the sample is far from being complete, it includes variety of objects.
There are three Hickson's compact groups, 
another three (NGC 1132, NGC 1550 and NGC 6521) 
X-ray selected groups discovered by the ROSAT survey,
and 11 relatively loose groups.
Their velocity dispersion ranges from $\sigma = 169$ to $474$ km s$^{-1}$.

\begin{longtable}{lrlllllll}
\caption{The sample objects selected for the present study.}
\label{tbl:obs:tarlist}
\hline\hline
target  & \multicolumn{2}{c}{position \footnotemark[$*$] } & $D$ \footnotemark[$\dagger$] & z \footnotemark[$\ddagger$] & $\sigma$ \footnotemark[$\S$] & $L_{\rm B}$ \footnotemark[$\|$] & $N_{\rm H}$ \footnotemark[$\#$] & another name\\
        & \multicolumn{2}{c}{( $\alpha$, $\delta$ )}&  Mpc    &       & km s$^{-1}$  &  $10^{11} L_{\odot}$   &  $10^{20}$ cm$^{-2}$     & \\
\hline
\endhead        

\hline
\endfoot
\hline
\multicolumn{9}{@{}l@{}}{\hbox to 0pt{\parbox{160mm}{\footnotesize
     \par\noindent
     \footnotemark[$*$] Position of the 0.5-10 keV X-ray centroid determined from the averaged image from GIS2 and GIS3.
     \par\noindent
     \footnotemark[$\dagger$] Distance to the group, converted from the recession velosity listed in the NED database, assuming a Hubble constant of $H_0 = 75$ km s$^{-1}$ Mpc$^{-1}$.
     \par\noindent
     \footnotemark[$\ddagger$] Redshift, from NED database.
          \par\noindent
     \footnotemark[$\S$] Radial velosity dispersion. Groups without index are from Zabludoff, Mulchaey (1998),
``$h$'' from Hickson (1982),``$d$'' from Drinkwater, Gregg, Colless (2001), ``$fs$'' from Ferguson, Sandage (1990), ``$w$'' from Wegner, Haynes, Giovanelli (1993),
``$f$'' from Fadda et al.~(1996), ``$rc$'' from RC3 catalog  (de Vaucouleurs et al.~1991),
``$r$'' from Ramella et al.~(1995), and ``$l$'' from Ledlow et al.~(1996).
     \par\noindent
     \footnotemark[$\|$] Optical B-band luminosity summed over the member galaxies of the group,
obtained from RC3 catalog.
     \par\noindent
     \footnotemark[$\#$] Galactic absorption column density derived from HI radio emission map by Dickey, Lockmann (1990).
     }\hss}}

\endlastfoot
HCG 51   &  $170.614$,&$24.294$         & 103   &0.0258 &240$^h$        & 1.05   & 1.27  &\\    
HCG 62   &  $193.277$, &$-9.209$        & 58.4  &0.0146 &$376^{+52}_{-46}$& 0.6 & 3.01&\\ 
HCG 97   &  $356.844$,&$-2.303$         & 87.2  &0.0218 &372$^h$        & 0.62    & 3.65  &\\   
NGC 507  &  $20.901$,&$33.257$          & 65.8  &0.0165 &595$^w$        & 1.73    & 5.24  &\\ 
NGC 533  &  $21.397$,&$1.772$           & 73.7  &0.0184 &$464^{+58}_{-52}$& 0.86  & 3.10  &\\
NGC 1132 &  $43.223$,&$-1.275$          & 92.8  &0.0232 & --            & 0.47  & 5.17  & \\  
NGC 1399 &  $54.622$,&$-35.450$         & 19.0  &0.0048 &374$^{d}$     & 0.45    & 1.34  & Fornax cluster\\ 
NGC 1550 &  $64.908$,&$2.410$           & 49.6  &0.0123 & --            & 0.21  & 11.5  & RX J0419.6+0225\\ 
NGC 2563 &  $125.149$,&$21.068$         & 59.8  &0.0149 &$336^{+44}_{-40}$& 0.55 & 4.23  &\\ 
NGC 4325 &  $185.796$,&$10.606$         & 102   &0.0257 &$265^{+50}_{-44}$& 0.48 & 2.22  &\\ 
NGC 5044 &  $198.859$,&$-16.398$        & 36.1  &0.0090 &474$^{fs}$     & 0.73   & 4.93  & WP 23\\  
NGC 5846 &  $226.622$,&$1.606$          & 24.3  &0.0061 &$368^{+72}_{-61}$& 0.57  & 4.26  &\\   
NGC 6329 &  $258.562$,&$43.684$         & 110   &0.0276 & --            & 1.03   & 2.12  &\\ 
NGC 6521 &  $268.942$,&$62.604$         & 106   &0.0266 &387$^z$        & 0.67   & 3.39  & RX J1755.8+6236\\ 
NGC 7619 &  $350.060$,&$8.206$          & 50.1  &0.0125 &780$^f$        & 1.08  & 0.50  & Pegasus group\\ 
Pavo     &  $304.628$,&$-70.859$        & 56.0  &0.0137 &169$^{rc}$     & 1.71 & 7.00 &\\ 
RGH 80   &  $200.058$,&$33.146$         & 148   &0.0370 &467$^r$        & 0.32    & 1.05  & USGC U530\\ 
S49-147  &  $  5.375$,&$22.402$         & 76.0  &0.0190 &$464^{+59~l}_{-21}$& 1.13 & 4.06 &\\ 
\end{longtable}

\begin{longtable}{lllll}
\caption{Log of the ASCA observations utilized in the present work.}
\label{tbl:obs:obslog}
\hline \hline
target  & r\footnotemark[$*$]  & sequence ID\footnotemark[$\dagger$] (year)    & \multicolumn{2}{c}{exposure \footnotemark[$\ddagger$] } \\
	&	&				& GIS	& SIS\\
\hline
\endhead
\hline
\endfoot
\hline
\multicolumn{5}{@{}l@{}}{\hbox to 0pt{\parbox{180mm}{\footnotesize
     \par\noindent
     \footnotemark[$*$] Radius of the spectral integration region. 
          \par\noindent
     \footnotemark[$\dagger$] Observation IDs. The SIS data are extracted only from the observations with ``$s$''.
     Associated number represents the CCD mode, such as 1, 2 and 4 CCD modes.
     Offset pointing are labeled as ``$o$''.
          \par\noindent
     \footnotemark[$\ddagger$] The total effective exposure of the observation, including the offset pointings.
               \par\noindent
     \footnotemark[$\S$] The data only from the SIS0 is used. The shape of the SIS1 spectra was odd, 
     and inconsistent with the short supplemental observation (ID=61007010). 
     \par\noindent
     \footnotemark[$\|$] The SIS data from the first observation is not used.
     The CCD temperature was too high for the 4 CCD mode operation and the shape of the spectra is
     severely distorted and unreliable  (e.g. Finogenov et a. 2002).
     }\hss}}
\endlastfoot
HCG 51  &$10'$& 82028000('94)$^{s2}$       & 62          & 72  \\
HCG 62	&$15'$& 81012000('94)$^{s2}$, 86008000('98),86008010('98), 86008020('98),86008020('98) & 121 & 29 \\
HCG 97  &$10'$& 84006000('96)$^{s1}$      & 79        &81    \\
NGC 507 &$15'$& 61007000('94)$^{s2}$\footnotemark[$\S$] , 61007010('95), 63026000('95)$^o$     & 80     &25 \\
NGC 533 &$10'$& 62009000('94) \footnotemark[$\|$] , 62009010('96)$^{s2}$ & 35    &18\\
NGC 1132 &$10'$& 65021000('97)$^{s1}$     & 27        &20    \\ 
NGC 1550 &$15'$& 87005000('99)$^{s1}$	& 70    &23\\ 
NGC 1399 &$20'$& 80038000('93)$^{s4}$, 80039000('93), 81021000('94)$^o$, 87006000('99)$^o$, & 145    &17\\
	& & 87006010('99)$^o$, 87006020('99)$^o$,87006030('99)$^o$, 87006040('99)$^o$ & &\\
NGC 2563 &$12'$& 63008000('95)$^{s1}$     & 46	 &52\\
NGC 4325 &$10'$& 85066000('97)$^{s2}$     & 27 	&25\\ 
NGC 5044 &$15'$& 80026000-10('93)$^{s4}$, 87002000('99)$^o$, 87002010('99)$^o$, 87002020('99)$^o$, 87002030('99)$^o$ & 111 &19   \\
NGC 5846 &$15'$& 61012000('94)$^{s4}$     & 36     &28\\
NGC 6329 &$12'$& 84047000('96)$^{s2}$     & 37     &34\\
NGC 6521 &$10'$& 85034000('97)$^{s1}$     & 36     &19\\ 
NGC 7619 &$12'$& 63017000('95)$^{s2}$     & 56     &59\\
Pavo    &$10'$& 81020000('94)$^{s4}$      & 29     &26\\
RGH 80  &$10'$& 83012000('95)$^{s2}$, 93007040('95)$^o$,93007080('95)$^o$, 93007070('95)$^o$       & 67  &42\\
S49-147 &$15'$& 81001000('93)$^{s4}$      & 32     &29\\
\end{longtable}

Table \ref{tbl:obs:obslog} summarizes ASCA observation log of
the objects in our sample. 
The 18 groups are covered by 39 observations in total, 
of which about one third are offset pointings.
We analyzed the GIS data in all observations.
In contrast, the SIS data are utilized only in the earliest
observation of the relevant object, unless otherwise noted.
This is to avoid the significant changes of the SIS response with time (e.g. Hwang et al.~1999).

\section{Data screening and background estimation}
\label{chap:screen}

Our strategy is to use the GIS and SIS in combination, making the best use of their characteristics.
The GIS has a higher stopping power,
together with a lower and stabler detector non X-ray background (NXB).
In addition, the background of the GIS has been extensively studied,
including both the Cosmic X-ray background (CXB) and the NXB
(Ishisaki 1997, Kushino et al.~2002).
Thus, the GIS data provide the principal probe with which to search for 
hard emission.
However, the GIS is not as good as the SIS in diagnostics of soft thermal X-ray spectra,
because of a lower energy resolution and a poorer low-energy quantum efficiency than those of the SIS.
Therefore, we employ the SIS data in order to accurately fix the thermal IGM emission.

\begin{figure}[htbp]
	\begin{center}
	\FigureFile(80mm,50mm){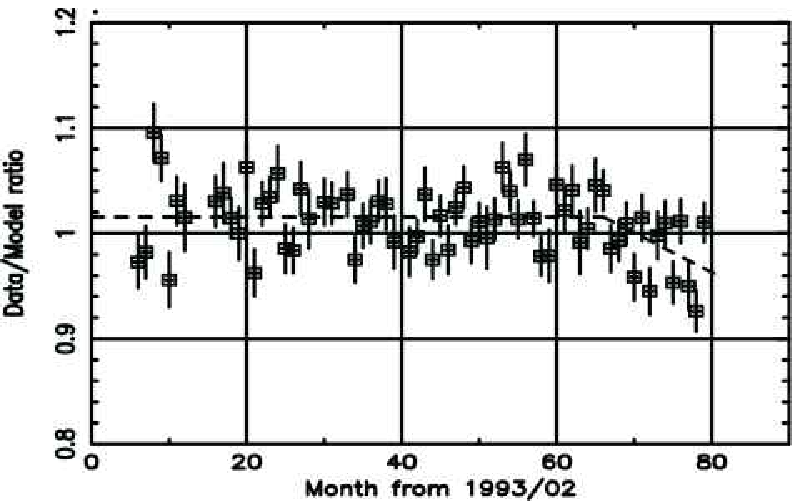}
	\FigureFile(80mm,50mm){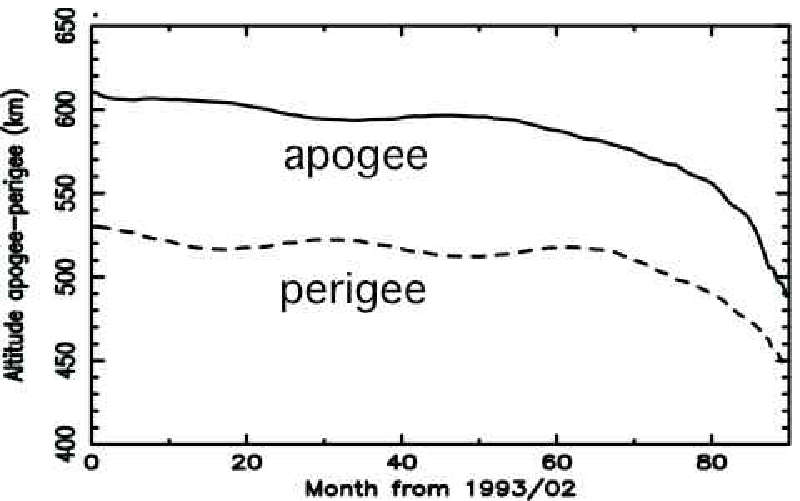}
	\end{center}
   	\caption{(left) The 4-8 keV GIS counting rate in the $r<15'$ region averaged over each month,
	normalized to the background modeled with our method. See text for detail.
	The dashed line is an analytic model fit.  ~(right) Changes of the apogee and perigee heights of ASCA.  }
\label{fig:screen:nxberr}
\end{figure}

The data from SIS0 and SIS1 were screened with ``{\it rev.2 standard processing}'',
which selects those data taken under the cut-off rigidity higher then 6 GV,
and with the source angle above the bright and night earth rim higher than
20$^{\circ}$ and 10$^{\circ}$, respectively. 
We utilized the bright mode data and coadded the 
two (SIS0 and SIS1) spectra into a single SIS data.
The background spectra were obtained from blank-sky observations,
filtered through the same procedure.

Similarly, 
we added the data from the two GIS detectors (GIS2 and GIS3) of each observation, 
and further added (if multiple pointings) all observations 
to obtain an average GIS spectrum of each target.
In order to subtract the background with the highest accuracy,
in this paper we adopted the procedure used by Ishisaki (1997), 
rather than simply using the data from ``{\it rev.2 standard}'' screening.
This method, called ``H02 method", not only employs 
a tighter set of data screening conditions,
but also models the GIS NXB according to
the distribution of counting rates of the events 
rejected in the on-board anti-coincidence circuit.
It allows us an NXB estimation with a 1$\sigma$ systematic error 
of 6\% and 3\%, for 10 ks and 40 ks observations, respectively.

To estimate the NXB in each on-source GIS dataset,
the H02 method needs NXB templates.
For this purpose, we prepared a data base
consisting of all the GIS events detected
when the ASCA telescope was pointing to the night Earth,
over a period from 1993 July through 2000 July.
The total exposure amounts to 5.2 Ms.
All the events in the data base
were sorted into 8 subsets, according to instantaneous values of ``H02 scalar''
which counts the GIS anti-coincidence rate.
Each subset then defines a GIS background template,
and the actual template to be
subtracted from a particular on-source spectrum is
constructed as a weighted mean of these templates.
Combined with this method, 
we estimated the CXB using data from 4 different blank sky regions
at high galactic latitudes with $|b| > 29^{\circ}$,
after eliminating discrete sources brighter than $\sim 2\times 10^{-13}$ erg s$^{-1}$ cm$^{-2}$
in the 2--10 keV band.

In our analysis, the NXB spectra around 4-8 keV is very important.
In order to further improve the accuracy of the NXB estimation,
we utilize the 5.9--10.6 keV counts 
in the GIS detector periphery ($15'-22'$ from the detector center),
hereafter denoted $N_{\rm out}^{\rm hard}$,
to adjust residual systematic differences in the estimated and actual NXB.
In calculating $N_{\rm out}^{\rm hard}$, we excluded a region
of $3'$ radius around each bright point source, 
and corrected $N_{\rm out}^{\rm hard}$ for the  decrease of the detector area.
About 80\% of $N_{\rm out}^{\rm hard}$ is due to the NXB,
while the remaining 20\% the CXB.
Any IGM signal is considered to contribute $\leq 0.1$ \% to $N_{\rm out}^{\rm hard}$
assuming typical spectral and spatial parameters of the IGM emission.
Then, we calculated ``correction factor",
as a ratio of $N_{\rm out}^{\rm hard}$ between the on-source 
data, and the background model in which
the CXB contribution is fixed to those from the blank sky observations.
Since a total of 80 ks exposure results in about $N_{\rm out}^{\rm hard}$ = 2000 counts,
we can estimate this factor with a statistical accuracy of 
$1/\sqrt(2000)/0.8 = 2.8$\% ($1\sigma$).

Using the night-earth event data base
utilized to construct the NXB templates,
we evaluated the accuracy with which the NXB can be reproduced.
The entire events in the same data base were again sorted into subsets,
but this time, with each subset covering one month.
The monthly exposure scatters between 18 ks and 150 ks,
with an average of 85 ks; those months with the exposure less than 40 ks were discarded.
From each monthly-accumulated night Earth spectrum,
we subtracted the template background synthesized
in the same way as before using the H02 method 
with the correction using $N_{\rm out}^{\rm hard}$.
(Each monthly spectrum and the template are both
derived from the same data base, but using different sortings.)
Figure \ref{fig:screen:nxberr} shows
the ratio of the count rate of each monthly spectrum,
to that dictated by the synthesized template.
While the $N_{\rm out}^{\rm hard}$ correction utilizes the 5.9--10.6 keV band,
the comparison in figure \ref{fig:screen:nxberr} (left)
is carried out in the 4--8 keV band,
which is used in the following sections to search for excess hard X-ray emission.
Including the statistical error of the 4-8 keV counts, 
our method provides an rms scatter of 3.3\%,
which is better than what is achieved by the original H02 method alone (4.0\%).

In  figure \ref{fig:screen:nxberr} (left), 
the ratio exhibits a gradual decrease  by about 1-2\%,
after the month 66 (1999 July);
this is due to orbit decay of ASCA,
as is clear from  figure \ref{fig:screen:nxberr} (right).
By representing this trend by two linear segments
as shown by a dashed line  in the figure,
the reproducibility of our NXB estimation method
was improved to 3.0\%.
In the following analysis, 
the systematic error in the NXB estimation in each target
is defined as a quadrature sum of the statistical error
associated with $N_{\rm out}^{\rm hard}$,
and a 1\% systematic error representing any residual unknown factors.

\section{Data Analysis and Results}
\label{chap:spec}

\subsection{X-ray images}
\label{chap:spec:overview}

\begin{figure*}[htbp]
	\begin{center}
	\FigureFile(130mm,150mm){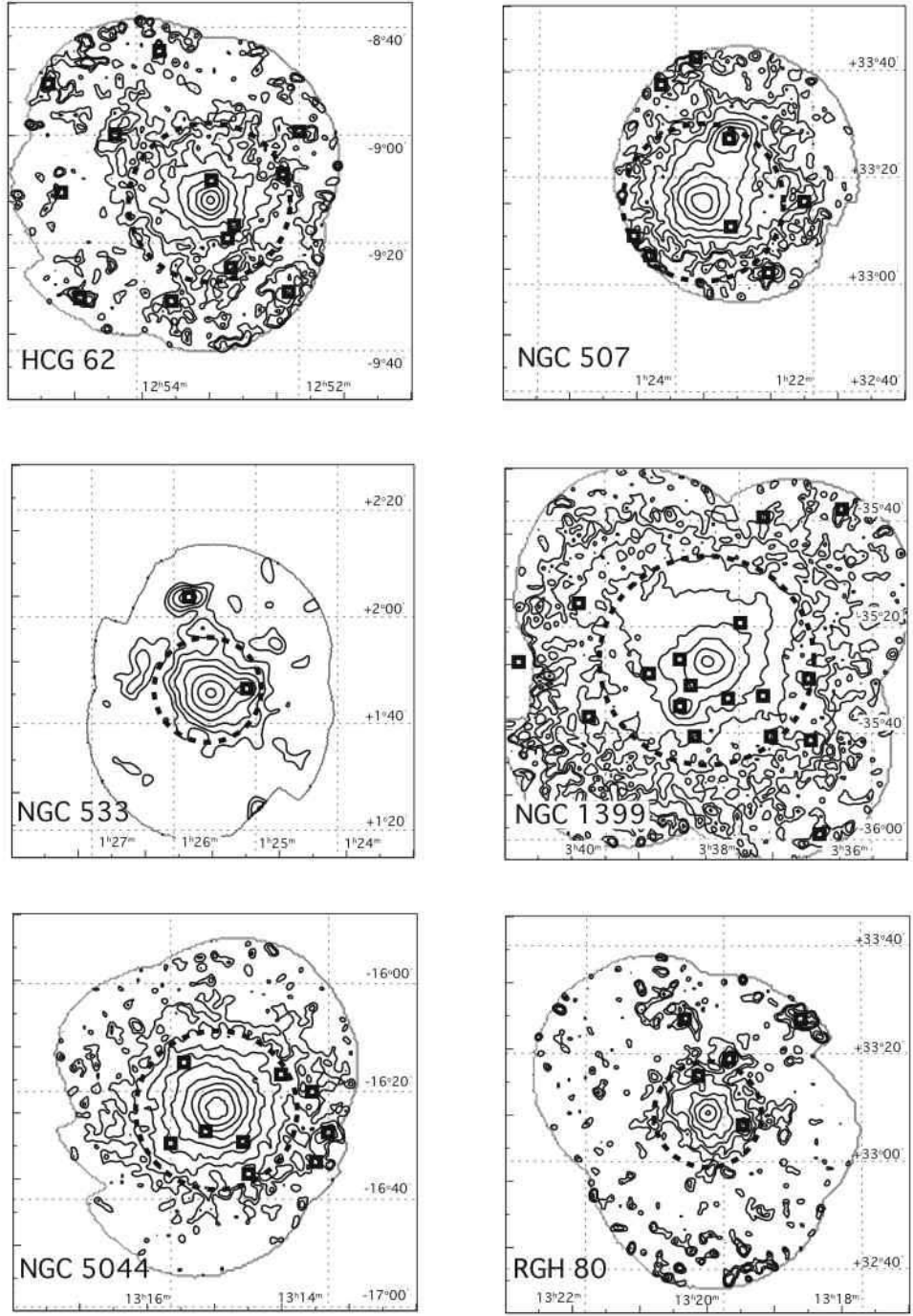}
	\end{center}
\caption{Background-subtracted  0.5--10 keV GIS (GIS2 plus GIS3) 
mosaic images of the groups of galaxies in the present sample,
observed under multiple pointings.
Each image is corrected for overlapping exposure,
and is  smoothed by a Gaussian function with $\sigma = 1'$.
Contours are logarithmetically spaced with a factor of 1.7, 
starting from $3\times 10^{-5}$ cts s$^{-1}$ cm$^{-2}$ arcmin$^{-2}$.
In NGC 1399 and NGC 5044, the least significant contour is deleted for simplicity.
The gray thin line indicates the combined GIS field of view.
Dashed circle shows the region used for spectral accumulation.
Boxes represent point sources eliminated in the spectral analysis.
}
\label{fig:other:image1}
\end{figure*}

\begin{figure*}[htbp]
	\begin{center}
	\FigureFile(160mm,180mm){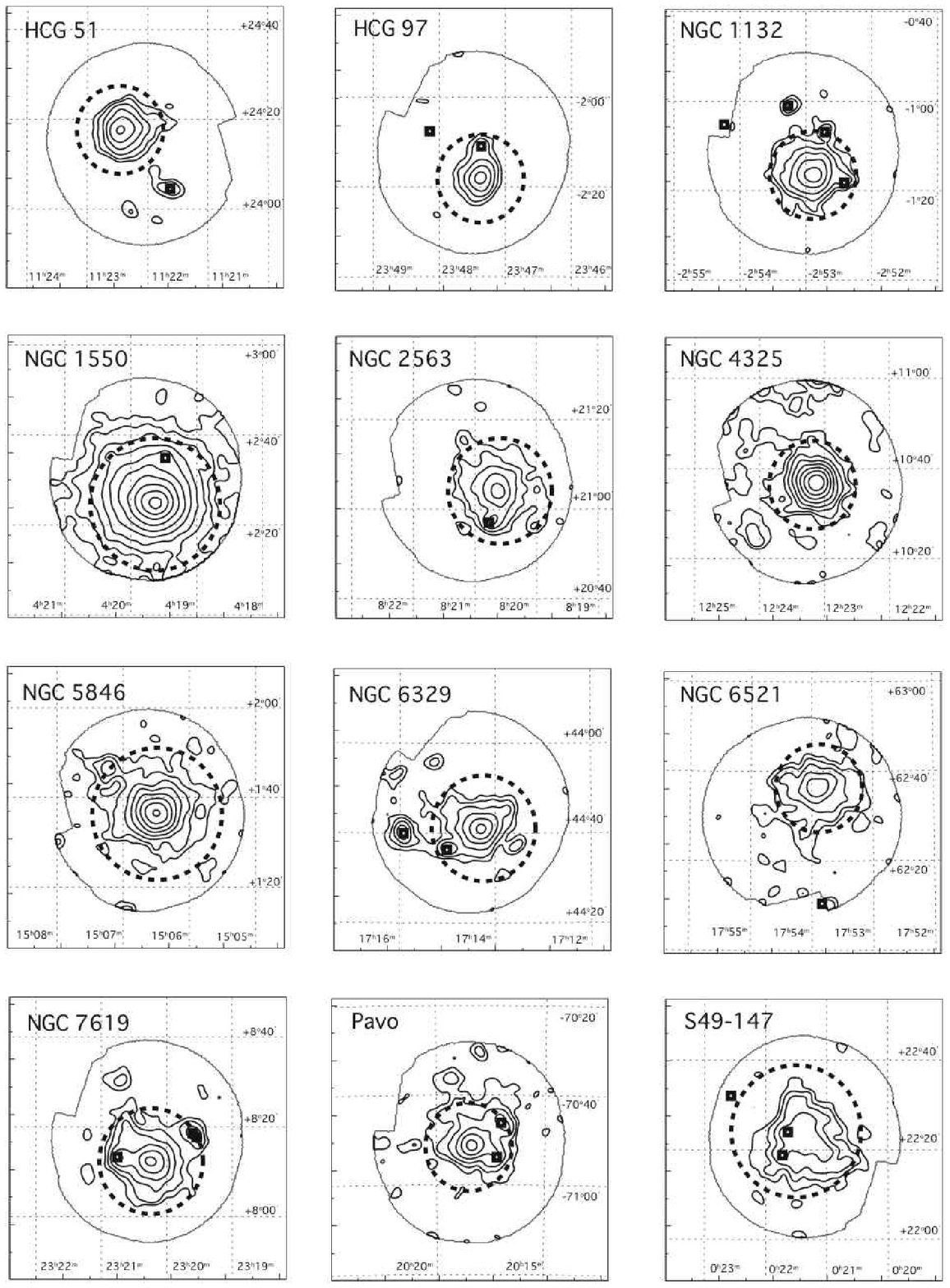}
	\end{center}
\caption{The same as figure~\ref{fig:other:image1}, but for the groups with single pointing.}
\label{fig:other:image2}
\end{figure*}

Figure~\ref{fig:other:image1} and figure~\ref{fig:other:image2}
show 0.5--10 keV GIS images of the 18  objects, 
obtained after subtracting the background (NXB and CXB) as derived  above.
Every object thus exhibits diffuse IGM emission with a roughly circular profile,
which is detected up to a radius of $10' - 25'$.
The X-ray centroid is generally coincident in position 
with a bright (often the brightest) elliptical galaxy in the system.
For each object, we then defined a spectral integration region,
as indicated in these figures (dashed circles) and listed in table \ref{tbl:obs:obslog}.
These regions are fully covered by the GIS, but only partially by the SIS.
We eliminated regions around bright contaminating sources,
such as the NGC 1404 galaxy close to NGC 1399 and the NGC 499 galaxy near NGC 507.
Similarly, we removed regions of $3'$ radius around point sources
in the 2RXP catalog\footnote{http://wave.xray.mpe.mpg.de/rosat/rra/rospspc},
if their 0.1--2.5 keV  count rates are higher than $4\times 10^{-3}$ counts s$^{-1}$.
From independent analysis using our GIS data, 
the 2--10 keV fluxes of these sources are confirmed to be
$\lesssim~2\times10^{-13}$ erg s$^{-1}$ cm$^{-2}$.
Positions of all removed sources are also presented in figure~\ref{fig:other:image2}.

\subsection{Basic characteristics of the X-ray spectra}
\label{chap:spec:basic}

In this section, we 
present the GIS and SIS spectra of our sample objects,
and quantify them through spectral model fittings.
A particular care is needed here,
because the soft part and hard part 
(typically below and above $\sim 2.5$ keV, respectively)
of the spectra are subject to distinct sources of errors.
Obviously, the hard-band data are strongly affected 
by the statistical and background uncertainties.
In contrast, the soft-band spectra have such high signal statistics
that  their uncertainties  are dominated by those in the instrumental responses
rather than  in the background.
Given these, 
we  first quantify the IGM emission using the soft-band spectra,
and then examine whether the results can explain the hard-band data or not.

\subsubsection{Fitting procedure}
\label{chap:spec:proc}

Figure \ref{fig:other:spec:single1} shows
background-subtracted GIS (black) and SIS (red) spectra of our 18 targets,
derived from the spectral accumulation regions defined above.
The GIS background has been subtracted
as detailed in section~\ref{chap:screen},
while the SIS background using 
blank-sky observations (also section ~\ref{chap:screen}).
Errors associated with the GIS data points include
the systematic background uncertainties noted in section~\ref{chap:screen}.
In contrast,  those assigned to the SIS spectra are statistical only,
because in this case Poisson errors 
of the signal and background counts are dominant,
in softer and harder energies, respectively.
The SIS spectra thus reveal Fe-L, Mg-K, and Si-K lines,
indicating the dominance of  thermal IGM emission,
whereas the GIS signals are generally
detectable to energies beyond $\sim 5$ keV.

Below,  we apply various spectral models 
simultaneously to the GIS and SIS spectra of each object,
utilizing energies above  1.0 keV and 0.7 keV, respectively.
To handle the difference between the GIS and SIS fields of view,
the model normalization
(of individual components if  the model is composite)
is allowed to differ between the two instruments.
When dealing with optically-thin thermal emission models,
we classify major heavy elements into two groups
in view of their origin (e.g., Matsushita 1997);
one group comprises  O, Ne, Na, Mg, Al, Si, S, Ar and Ca,
which are mainly so-called $\alpha$-elements,
while the other group  consists of Fe and Ni.
Abundances of the first and second metal groups 
are denoted as $Z_{\alpha}$ and $Z_{\rm Fe}$, respectively.
For the IGM emission calculations, 
we used both the vMEKAL and vAPEC codes 
provided by the XSPEC package.
Since the two models give only minor differences, 
below we refer only to the vMEKAL results.
The absorption column density was fixed to the value 
derived from HI observations (Dickey, Lockman, 1990).
These parameters are also listed in table \ref{tbl:obs:tarlist}.

The SIS and GIS responses have some residual uncertainties,
which are dominated by their gain calibration errors 
as a function of the detector position;
we quote those of the SIS and GIS as 
$\sim 0.5$\% and $\sim 1$\%, 
respectively\footnote{http://heasarc.gsfc.nasa.gov/docs/asca/cal\_probs.html}.
In some of the brightest objects in our sample, 
however, these values turned out to be insufficient to 
fully represent the instrumental calibration uncertainties
in the soft energy band where the signals have very high statistics.
As a conventional way to solve this problem,
we  allowed the model red-shift parameter to
vary independently within $\pm$ 0.5\% and $\pm$ 1\%, 
for the SIS and the GIS, respectively,
from that taken from the NED data base\footnote{http://nedwww.ipac.caltech.edu/}.
When analyzing the SIS data taken in the 4-CCD mode, 
the SIS gain tolerance was slightly relaxed to $\pm$ 1\%,
to incorporate additional gain uncertainties.

For the NGC 1550 group, we set  free the absorption column of the SIS data,
because the earliest data of this group was obtained in 1999 
when the SIS degradation had already been significant;
the additional absorption is expected to  
emulate changes in the SIS response (e.g. Hwang et al.~1999).
For the NGC 6329 group, we limited our analysis to energies above 0.85 keV for the SIS.
This is because a significant soft hump is observed below this energy, which is not
detected by, e.g., ROSAT (e.g. Mukchaey et al.~2003) and supposed to be instrumental.
Even if the humps is included in the fit, it requires very soft ($kT<0.1$ keV) component, 
which does not affect the results in the following sections.

\subsubsection{Single-temperature fits to the soft-band spectra}
\label{chap:spec:single}

\begin{figure*}[htbp]
	\begin{center}
	\FigureFile(160mm,160mm){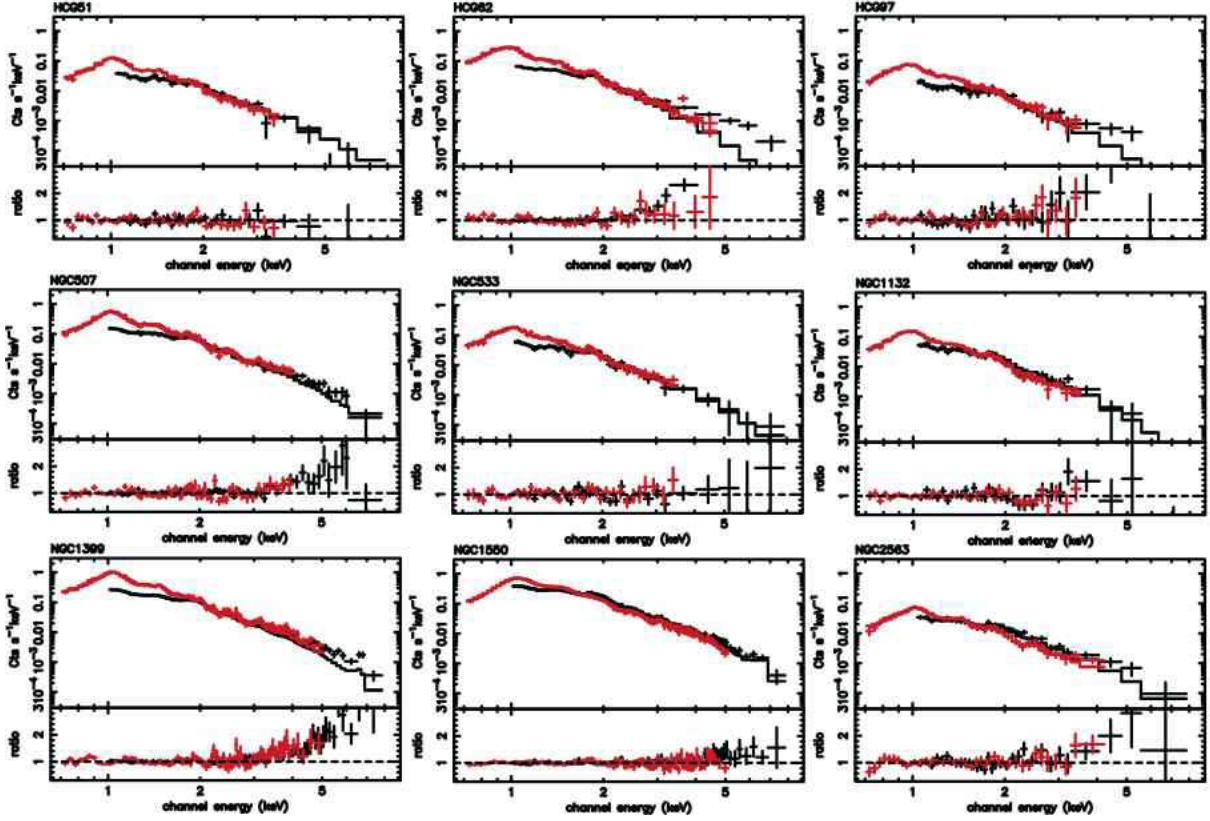}
	\end{center}
\caption{The GIS (black) and SIS (red) spectra of our sample objects,
jointly fitted with a vMEKAL model in the energy below 2.5 keV.
The models are extrapolated to higher energies. 
All spectra are plotted to the same scale.
}
\label{fig:other:spec:single1}
\end{figure*}

\addtocounter{figure}{-1}

\begin{figure*}[htbp]
	\begin{center}
	\FigureFile(160mm,160mm){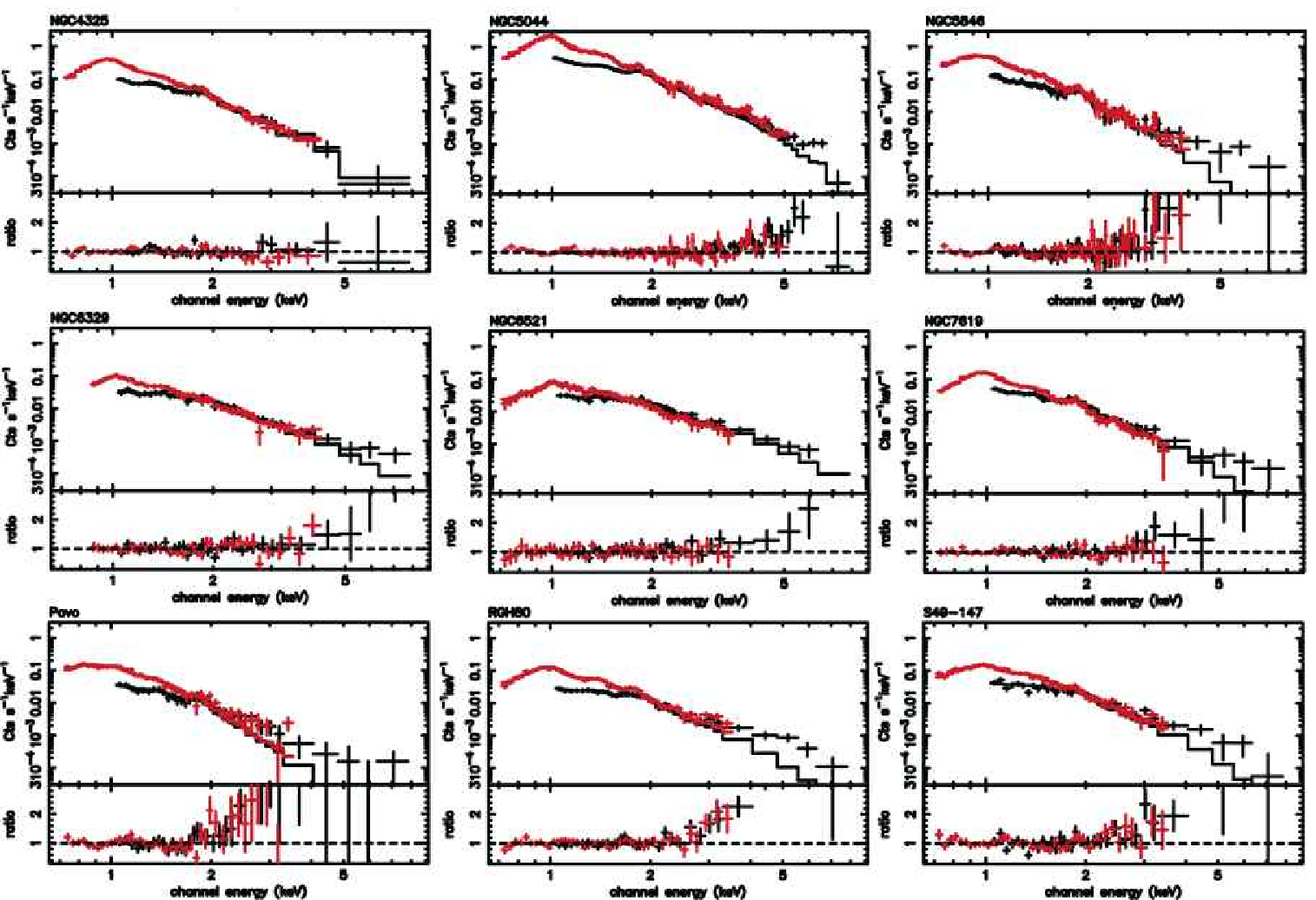}
	\end{center}
\caption{(continued)}
\end{figure*}

As the first attempt, 
we applied a common vMEKAL model 
to the GIS and SIS spectra of each object,
over a limited soft energy range below 2.5 keV.
The results are summarized in table \ref{tbl:other:spec:single}, 
and the obtained best-fit models are shown as a histogram in figure \ref{fig:other:spec:single1}.
There,  the models determined in the energy range below 2.5 keV
are extrapolated to higher energies,
to be compared with the observed data.
These soft-band fits gave the IGM temperature in the range $kT_{\rm S} =0.7-1.7$ keV,
together with sub-solar metal abundances.
The 0.7--2.5 keV band flux, $F_{\rm soft}$, 
was obtained in the range of
1 -- 20 $\times 10^{-12}$ erg s$^{-1}$ cm$^{-2}$.

In nearly half the objects,
the single-temperature fits are  in fact formally unacceptable.
In NGC~507, NGC~1399, NGC~1550, NGC~5044, NGC~5846, and Pavo,
the fit leaves significant residuals around atomic emission lines.
This suggest that their spatially-integrated IGM emission 
cannot be described adequately with single-temperature thermal models:
later in section~4.2.4, we hence introduce two-temperature models. 

In figure \ref{fig:other:spec:single1}, 
we also notice that the data at energies above $\sim 3$ keV
often exceed the model extrapolation,
in at least 6 objects such as
HCG 62, NGC 507, NGC 1399, and RGH 80.
This suggests the presence of additional harder emission components in these objects.
Actually, if we fit the GIS/SIS spectra of these objects over 
the whole energy band including data points above 2.5 keV,
the reduced chi-squared further increases by $\sim$0.3 or more.

\begin{longtable}{l|lllrlll}
\caption{Results of the joint fit to the GIS and SIS spectra below 2.5 keV
with a single component vMEKAL model. }
\label{tbl:other:spec:single}
\hline \hline
target  & $kT_{\rm S}$\footnotemark[$*$] & \multicolumn{2}{c}{Abundance \footnotemark[$\dagger$] }             
& $\chi^2$/dof & $F_{\rm soft}$ \footnotemark[$\ddagger$]  & $kT_{\rm H}$ \footnotemark[$\S$] & note \footnotemark[$\|$] \\
            & (keV)          & $Z_{\alpha}$($Z_{\odot}$) & $Z_{\rm Fe}$($Z_{\odot}$) & &     &             & \\
\hline
\endhead
\hline
\endfoot
\hline
\multicolumn{9}{@{}l@{}}{\hbox to 0pt{\parbox{140mm}{\footnotesize
     \par\noindent
     \footnotemark[$*$] Best fit temperature in the energy band below 2.5 keV.
          \par\noindent
     \footnotemark[$\dagger$] Best fit metal abundances in the energy band below 2.5 keV.
          \par\noindent
     \footnotemark[$\ddagger$] Flux in the energy band from 0.7 to 2.5 keV ($10^{-12}$ erg s$^{-1}$ cm$^{-2}$).
               \par\noindent
     \footnotemark[$\S$] Best fit temperature from the hard-band fitting, above 2.5 keV. Errors are the 90\% statistical and systematic ones.
               \par\noindent
     \footnotemark[$\|$] Targets in which the soft-band fitting is not statistically acceptable in 90\% and 99\% confidence are indicated with "$\star$" and "$\star\star$, respectively.
     }\hss}}
\endlastfoot
HCG 51&$1.37_{-0.03}^{+0.03}$&$0.44_{-0.12}^{+0.15}$&$0.36_{-0.05}^{+0.07}$&72.0/52	&1.9	&$0.96_{-0.31}^{+0.50}$$_{-0.41}^{+0.74}$&$\star$\\
HCG 62&$0.95_{-0.04}^{+0.03}$&$0.31_{-0.07}^{+0.09}$&$0.15_{-0.03}^{+0.03}$&70.0/52	&4.0	&$2.47_{-0.46}^{+0.61}$$_{-0.57}^{+0.55}$&$\star$\\
HCG 97&$1.03_{-0.06}^{+0.04}$&$0.31_{-0.12}^{+0.16}$&$0.19_{-0.04}^{+0.05}$&59.5/52	&0.9	&$1.88_{-0.84}^{+1.74}$$_{-1.15}^{+1.29}$&\\
NGC 507&$1.35_{-0.03}^{+0.03}$&$0.62_{-0.10}^{+0.12}$&$0.41_{-0.05}^{+0.06}$&128.8/96	&7.8	&$1.86_{-0.19}^{+0.19}$$_{-0.19}^{+0.17}$&$\star$\\
NGC 533&$1.23_{-0.10}^{+0.08}$&$0.52_{-0.18}^{+0.26}$&$0.31_{-0.08}^{+0.11}$&80.7/52	&2.7	&$1.30_{-0.46}^{+0.65}$$_{-0.57}^{+0.66}$&$\star\star$\\
NGC 1132&$1.08_{-0.04}^{+0.04}$&$0.33_{-0.11}^{+0.15}$&$0.27_{-0.05}^{+0.06}$&53.9/52	&2.7	&$1.09_{-0.36}^{+0.51}$$_{-0.37}^{+0.48}$&\\
NGC 1399&$1.31_{-0.04}^{+0.02}$&$0.56_{-0.04}^{+0.03}$&$0.35_{-0.04}^{+0.03}$&239.9/104	&14.3	&$1.84_{-0.11}^{+0.09}$$_{-0.11}^{+0.09}$&$\star\star$\\
NGC 1550&$1.38_{-0.02}^{+0.02}$&$0.58_{-0.06}^{+0.07}$&$0.42_{-0.03}^{+0.04}$&191.9/103	&19.1	&$1.52_{-0.06}^{+0.07}$$_{-0.08}^{+0.09}$&$\star\star$\\
NGC 2563&$1.32_{-0.07}^{+0.10}$&$0.64_{-0.21}^{+0.31}$&$0.28_{-0.07}^{+0.14}$&45.7/52	&1.6	&$1.73_{-0.57}^{+0.94}$$_{-0.80}^{+0.84}$&\\
NGC 4325&$1.02_{-0.02}^{+0.02}$&$0.42_{-0.12}^{+0.16}$&$0.32_{-0.05}^{+0.07}$&61.0/52	&4.6	&$1.20_{-0.41}^{+0.60}$$_{-0.43}^{+0.59}$&\\
NGC 5044&$0.99_{-0.01}^{+0.01}$&$0.39_{-0.04}^{+0.04}$&$0.27_{-0.02}^{+0.02}$
&261.3/104  &23.4   &$1.36_{-0.11}^{+0.11}$$_{-0.12}^{+0.12}$&$\star\star$\\
NGC 5846&$0.67_{-0.02}^{+0.02}$&$0.40_{-0.11}^{+0.20}$&$0.18_{-0.03}^{+0.05}$&136.8/98	&7.0	&$2.18_{-0.89}^{+2.34}$$_{-1.21}^{+1.24}$&$\star\star$\\
NGC 6329&$1.56_{-0.15}^{+0.15}$&$0.63_{-0.29}^{+0.43}$&$0.39_{-0.17}^{+0.22}$&38.2/47	&1.6	&$2.42_{-0.86}^{+1.56}$$_{-1.00}^{+0.92}$&\\
NGC 6521&$1.74_{-0.29}^{+0.35}$&$0.69_{-0.33}^{+0.45}$&$0.38_{-0.18}^{+0.24}$&36.5/52	&1.6	&$1.85_{-0.45}^{+0.62}$$_{-0.49}^{+0.50}$&\\
NGC 7619&$0.95_{-0.04}^{+0.04}$&$0.42_{-0.13}^{+0.15}$&$0.23_{-0.05}^{+0.05}$&55.9/52	&2.7	&$1.69_{-0.63}^{+1.00}$$_{-0.92}^{+0.96}$&\\
Pavo&$0.54_{-0.05}^{+0.06}$&$0.17_{-0.10}^{+0.18}$&$0.06_{-0.03}^{+0.02}$&70.1/52	&1.9	&$0.73_{-0.43}^{+1.13}$$_{-0.36}^{+1.30}$&$\star$\\
RGH 80&$1.09_{-0.03}^{+0.08}$&$0.50_{-0.14}^{+0.17}$&$0.18_{-0.03}^{+0.04}$&50.9/52	&1.3	&$2.62_{-0.68}^{+1.09}$$_{-0.79}^{+0.77}$&\\
S49-147&$0.97_{-0.08}^{+0.08}$&$0.15_{-0.08}^{+0.09}$&$0.06_{-0.02}^{+0.02}$&81.7/52	&2.2	&$2.38_{-0.90}^{+1.94}$$_{-0.87}^{+0.83}$&$\star\star$\\
\end{longtable}

\begin{figure}[htbp]
	\begin{center}
	\FigureFile(60mm,60mm){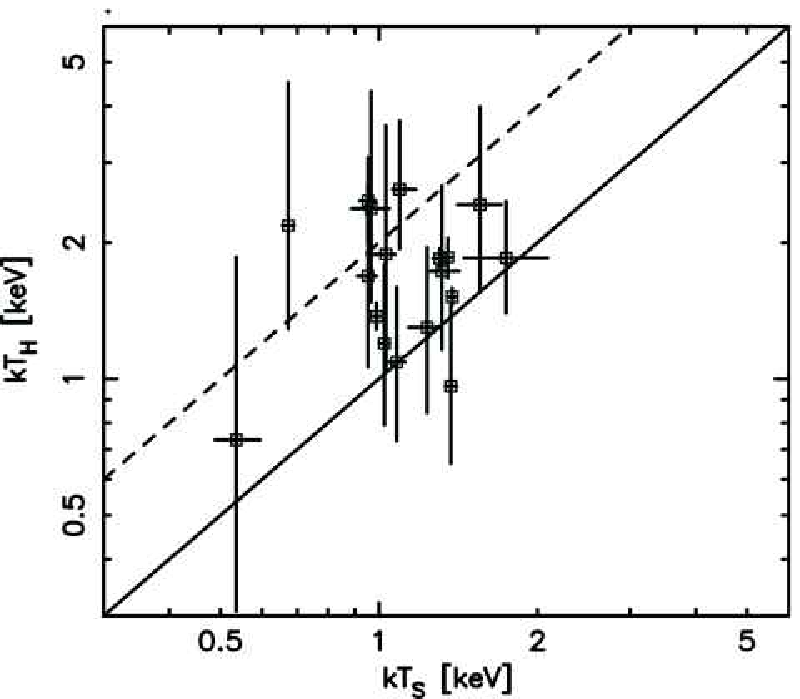}
	\end{center}
\caption{The best fit temperature from the hard-band fitting ($kT_{\rm H}$),
compared to that from the soft-band fitting ($kT_{\rm S}$).
Statistical 90\% errors are plotted.
The solid line indicates $kT_{\rm H}=kT_{\rm S}$, 
while the dashed line  $kT_{\rm H}=2kT_{\rm S}$.
}
\label{fig:other:anaspec:single:ktsh}
\end{figure}

\subsubsection{Properties of the hard-band spectra}
\label{chap:spec:kth}

In order to characterize the spectra in the hard band, 
we next analyzed  the 2.5--8 keV GIS  data of our sample objects.
The SIS data were not incorporated, 
since the SIS background systematics above 2.5 keV is not well understood.
We employed the same single-temperature vMEKAL model,
but fixed $Z_{\alpha}$ and $Z_{\rm Fe}$  both to 0.5 solar for simplicity,
because this energy range is relatively devoid of strong
metal lines when the plasma temperature is $\lesssim 2$ keV.
The red-shift is also fixed at the optical value.
Then, the temperature in this energy range, denoted $kT_{\rm H}$, 
is to be determined  by the continuum shape.

Since the GIS data above 2.5 keV are dominated by the backtround (NXB+CXB),
errors associated with  $kT_{\rm H}$ in this case 
are expected to come from three major sources;
photon counting statistics, 
systematic errors in the NXB subtraction, and those of the CXB.
The formal ``fitting error''  takes into account the first two factors,
because the NXB estimation errors are already included 
in the error bars of individual GIS data points (section 4.2.1).
However, the third factor is not.
Accordingly, we estimated the CXB contribution 
as described in the end of this subsubsection,
and present them in addition to the fitting errors for  $kT_{\rm H}$.

The hard-band temperatures $kT_{\rm H}$,
obtained in this way,  
are  listed in table \ref{tbl:other:spec:single},
and are compared in figure~\ref{fig:other:anaspec:single:ktsh}
with the  soft-band  temperature $kT_{\rm S}$. 
Although the errors (taking into account all the three sources) are large, 
we find that $kT_{\rm H}$ is generally higher than $kT_{\rm S}$,
even reaching $\sim 2kT_{\rm S}$ in some objects (e.g., HCG 62 and RGH 80).
This result reinforces the positive data excess toward higher energies,
which are suggested by figure~\ref{fig:other:spec:single1}.
For reference, the obtained $kT_{\rm H}$ does not change significantly
even if the 2.5--4.0 keV SIS data are included to the fit.

The third factor causing errors in $kT_{\rm H}$, 
namely sky-to-sky fluctuations in the CXB surface brightness  $I_{\rm CXB}$,
occurs as the number of faint objects that constitute the CXB fluctuates.
Assuming a simple  Euclidean log$N$-log$S$ relation as $N(>S) \propto S^{-1.5}$,
where $S$ is the source flux and $N(>S)$ is the number of sources
with fluxes higher than $S$,
the CXB brightness fluctuation $\sigma_{\rm CXB}$ is described as
$\sigma_{\rm CXB}/I_{\rm CXB} \propto \Omega_{\rm e} ^{-0.5}S_{\rm c}^{0.25}$.
Here,  $S_{\rm c}$ is the upper flux bound of the individually eliminated sources,
which is  $\sim 2\times 10^{-13}$ erg s$^{-1}$ cm$^{-2}$ (2--10 keV) in this analysis,
and $\Omega_{\rm e}$ is the effective solid angle of the detector.
For the ASCA GIS,  $\Omega_{\rm e}$ has been calculated by Ishisaki (1997),
by taking into account the ASCA vignetting function.
In the present case, a  typical data accumulation radius of 
$15'$ gives $\Omega_{\rm e} = 0.1$ deg$^2$. 
By scaling the HEAO-1 A2 result (Shafer, Fabian 1998) with these values,
we typically find that the 2.5--8 keV CXB  fluctuates 
from field to field  by 7.9\% (1$\sigma$)  in the present case.
Then, by fitting the GIS spectra with the CXB level artificially changed
(by a level corresponding to the 90\% range of its fluctuation),
we evaluated the error propagation from the CXB brightness to $kT_{\rm H}$.

\subsubsection{Two-temperature IGM modeling}
\label{chap:spec:2kt}

The single-temperature vMEKAL fit to the soft band spectra
ended up unacceptable in 10 objects (section \ref{chap:spec:single}):
before examining the hard excess suggested by the large difference 
between $kT_{\rm S}$ and $kT_{\rm H}$ in several objects,
we need to arrive at an acceptable IGM modeling of all objects.
These  fit failures are most likely  caused
by their slight deviation from the assumed isothermality,
because temperature gradients in the X-ray emitting plasma are
rather common among groups and clusters of galaxies.
Even in such a case, 
two temperatures are generally known to be sufficient 
to describe the integrated X-ray spectrum  (e.g., Fukazawa 1997).

We applied a two-temperature vMEKAL model (2-vMEKAL model)
to the GIS and SIS spectra of the 10 objects, again 
in the energy range below 2.5 keV.
The two vMEKAL components were allowed to take
free temperatures and normalizations,
but were constrained to have the same abundance parameters.
To avoid strong couplings between them,
we constrained the hotter temperature ($kT_2$) to stay between 0.8 and 25 keV;
the former is a typical cooler component temperature $kT_1$,
while the latter is 10 times the upper bound of the employed energy range.

\begin{longtable}{llllllll}
\caption{Results of the 2-vMEKAL model fitting in energies below 2.5 keV .}
\label{tbl:other:excess:2vmekal}
\hline \hline
target  & Model\footnotemark[$*$]  & $kT_1$\footnotemark[$\dagger$] &$kT_2$\footnotemark[$\ddagger$]  & Abundance   
& $\chi^2$/dof & note\footnotemark[$\S$]  & $-\Delta\chi^2$\footnotemark[$\|$]  \\
            & & (keV)    & (keV)  & $Z_{\rm Fe}$($Z_{\odot}$) & &  & \\
\hline
\endhead
\hline
\endfoot
\hline
\multicolumn{8}{@{}l@{}}{\hbox to 0pt{\parbox{140mm}{\footnotesize
     \par\noindent
     \footnotemark[$*$] Model ID. ``2Z'' means that the abudances are 
     grouped into two ($Z_{\alpha}$ and $Z_{\rm Fe}$). 
     ``6Z'' means models with Mg, Al, Si and S set free, in addition to ``2Z''. ``6Z-'' means after
     subtracting the two energy bands. See text.     
          \par\noindent
     \footnotemark[$\dagger$] Temperature of the cooler component. 
          \par\noindent
     \footnotemark[$\ddagger$] Temperature of the hotter component. 
     Parameters hitting the limit (0.8--25 keV) are presented with $>$ or $<$.
          \par\noindent
     \footnotemark[$\S$] Targets in which the soft-band fitting is not statistically acceptable in 90\% and 99\% confidence are indicated with ``$\star$'' and ``$\star\star$'', respectively.
    \par\noindent
     \footnotemark[$\|$] Improvement in $\chi^2$ by introducing the second thermal component. 
Presented with parentheses, if the improvement is not significant with F-test by 1\%.
     \par\noindent
          }\hss}}
\endlastfoot
HCG 51&6Z&$0.98_{-0.23}^{+0.15}$&$1.76_{-0.38}^{+0.51}$&$0.70_{-0.33}^{+0.60}$
&59.7/45&$\star$& (12.3) \\
HCG 62&2Z&$0.78_{-0.23}^{+0.07}$&$1.27_{<0.80}^{+0.33}$&$0.24_{-0.07}^{+0.08}$
&50.6/49 & & 19.4 \\
NGC 507&6Z&$0.71_{-0.29}^{+0.70}$&$1.36_{-0.07}^{>25.0}$&$0.46_{-0.11}^{+0.33}$
&109.7/86 & $\star$& (19.1) \\
NGC 533&6Z&$1.12_{-0.78}^{+0.11}$&$1.42_{<0.80}^{+1.52}$&$0.34_{-0.08}^{+0.16}$
&68.1/45 & $\star$ & (12.6) \\
NGC 1399&6Z-&$0.82_{-0.14}^{+0.07}$&$1.43_{-0.10}^{+0.17}$&$0.44_{-0.05}^{+0.14}$
&107.4/79& $\star$ & 33.5 \\
NGC 1550&6Z-&$0.95_{-0.35}^{+0.33}$&$1.44_{-0.10}^{+0.14}$&$0.52_{-0.14}^{+0.40}$
&105.9/78& $\star$ &  (15.9)\\
NGC 5044&6Z-&$0.67_{-0.11}^{+0.09}$&$1.11_{-0.05}^{+0.05}$&$0.35_{-0.02}^{+0.05}$
&116.7/79& $\star\star$ & 61.0 \\
NGC 5846&2Z&$0.63_{-0.02}^{+0.02}$&$1.28_{-0.45}^{+0.07}$&$0.36_{-0.07}^{+0.04}$
&92.3/92 & & 44.5 \\
Pavo&2Z&$0.39_{-0.07}^{+0.09}$&$1.56_{-0.24}^{+0.84}$&$0.58_{-0.37}^{+8.35}$
&40.9/49 & & 29.2 \\
S49-147&6Z&$0.60_{-0.60}^{+0.19}$&$1.15_{<0.80}^{+0.21}$&$0.09_{-0.05}^{+0.06}$
&54.3/45 & & 27.4\\
\end{longtable}

Results of the 2-vMEKAL fits 
are summarized in table \ref{tbl:other:excess:2vmekal}.
Spectra of HCG 62, NGC 5846 and Pavo
have been fitted successfully by the 2-vMEKAL model 
when the abundances are grouped as $Z_{\alpha}$ and $Z_{\rm Fe}$.
When the abundances of Mg, Al, Si, and S are set free, S49-147 gave acceptable fit.
With the same model,
HCG 51, NGC 507 and NGC 533 gave marginally (at 99\% confidence) 
acceptable fits,
although $kT_2$  is not well determined in the latter two objects.

The remaining 3 objects (NGC 1399, NGC 1550 and NGC 5044) 
have such high statistics around 1 keV that the fits remained still unacceptable 
even setting the Mg, Al, Si, S abundances free.
However, the fit itself has greatly improved by adding the second component,
implying that their spectra clearly prefer the 2-vMEKAL model to the vMEKAL model.
Adding a third emission component does not improve the fit significantly.
The fit failure in these objects is partially due to slight discrepancy between the SIS and the GIS data, 
at $\sim 1.2$ keV and $\sim 2.1$ keV, presumably due to calibration errors.
(The same features are often present in other observations,
but usually negligible compared to data statistics.)
We hence decided to simply ignore the energy bands of 1.15--1.25 keV and 2.1--2.2 keV.
Then the fit improved significantly, and became marginally acceptable by 99\% in
NGC 1399 and NGC 1550, while still unacceptable by 99.6\% in NGC 5044.

\subsection{Excess Hard X-ray Signals}
\label{chap:spec:excess:n_excess}

Having fixed the IGM emission using the soft-band data,
let us examine the significance of the suggested 
excess hard X-ray emission by counting excess  photons,
employing the energy band from 4 to 8 keV.
In this range, the GIS still has an effective area of $40-80$ cm$^2$,
but contribution from the IGM emission is expected to be very small.
In fact, 70-99 \% of the raw count rate in this band is from the background,
of which the CXB and the NXB have comparable shares.

In table~\ref{tbl:other:excess:excess},
the 3rd column represents the raw (background inclusive) 4--8 keV GIS 
count rate of each object accumulated
over the specified data integration region
(figure~\ref{fig:other:image1} and figure~\ref{fig:other:image2}),
together with the purely statistical 1$\sigma$ errors.
The 4th column shows the expected CXB contribution,
of which the errors refer to $1\sigma$ sky-to-sky fluctuation 
calculated in the same way as in section~4.2.3 
(e.g., 7.9\% for a region of $15'$ in radius).
The 5th column represents the estimated NXB count rate,
derived as described in section~\ref{chap:screen}.
Since the NXB templates have high  photon statistics (section~\ref{chap:screen}),
the quoted errors are 1$\sigma$ systematic,
which is typically 3.2 \% and 2.0\% 
for 40 ks and 120 ks observations, respectively.
The 6th column gives the IGM contribution to the 4--8 keV band,
calculated by extrapolating the soft-band determined IGM model.
We used the 1-vMEKAL fit if it is acceptable,
whereas the 2-MEKAL fit otherwise.
Although uncertainties in the IGM contribution have
composite origin, they are dominated by
statistical errors in the soft-band temperature determination,
because the error is amplified as the model prediction 
is extrapolated toward the higher 4--8 keV range.
We hence calculated how the errors associated with 
the soft-band temperature determinations
propagate into those in the  4--8 keV counts.
The original 90\% confidence errors were
convert into $1\sigma$ values assuming a Gaussian distribution.

\begin{longtable}{lcllllll} 
\caption{Excess rates in the 4-8 keV band with IGM contribution calculated below 2.5 keV.}
\label{tbl:other:excess:excess}
\hline \hline
target  &Model\footnotemark[$*$] &Data\footnotemark[$\dagger$]& CXB \footnotemark[$\ddagger$]  & NXB  \footnotemark[$\S$]  & IGM  \footnotemark[$\|$]   & Excess \footnotemark[$\#$]  & Sigma\footnotemark[$**$]\\
\hline
\endhead
\hline
\endfoot
\hline
\multicolumn{7}{@{}l@{}}{\hbox to 0pt{\parbox{150mm}{\footnotesize
     \par\noindent
     \footnotemark[$*$] IGM model ID. ``1T'' mains vMEKAL and ``2T'' means 2-vMEKAL models.
          \par\noindent
    \footnotemark[$\dagger$] Rates in $10^{-3}$ cts s$^{-1}$ GIS$^{-1}$. Errors are statistical 1 $\sigma$.
         \par\noindent
     \footnotemark[$\ddagger$] Rates in $10^{-3}$ cts s$^{-1}$ GIS$^{-1}$. Errors are from fluctuation, 1 $\sigma$.
          \par\noindent
     \footnotemark[$\S$] Rates in $10^{-3}$ cts s$^{-1}$ GIS$^{-1}$. Errors are systematic, 1 $\sigma$.
               \par\noindent
     \footnotemark[$\|$] Rates in $10^{-3}$ cts s$^{-1}$ GIS$^{-1}$. See text for the IGM error.
               \par\noindent
     \footnotemark[$\#$] Statistical $1\sigma$ and quadrature sum of all systematic $1\sigma$ errors.
                 \par\noindent
     \footnotemark[$**$] Significance of excess rates compared to the quadrature 
     sum of all $1\sigma$ errors. Value of NGC 5044 is presented with brackets,
     because the IGM model fit in the soft band is not acceptable even at 99\% confidence.
        }\hss}}
\endlastfoot
HCG 51&2T&$10.96\pm0.30$&$6.12\pm0.73$&$4.76\pm0.12$&$1.15\pm0.45$&
$1.07\pm0.30\pm0.87$&$-1.17$\\
HCG 62&2T&$20.26\pm0.29$&$8.24\pm0.66$&$9.21\pm0.18$&$0.81\pm0.15$&
$2.00\pm0.29\pm0.70$&2.64\\
HCG 97&1T&$11.46\pm0.27$&$5.50\pm0.66$&$5.48\pm0.13$&$0.17\pm0.01$&
$0.32\pm0.27\pm0.67$&0.44\\
NGC 507&2T&$20.91\pm0.36$&$7.42\pm0.59$&$8.97\pm0.21$&$3.07\pm1.54$&
$1.45\pm0.36\pm1.66$&0.85\\
NGC 533&2T&$12.01\pm0.42$&$5.76\pm0.69$&$5.28\pm0.17$&$0.55\pm0.09$&
$0.42\pm0.42\pm0.72$&0.51\\
NGC 1132&1T&$10.21\pm0.43$&$5.64\pm0.68$&$4.53\pm0.17$&$0.52\pm0.03$&
$-0.48\pm0.43\pm0.70$&$-0.58$\\
NGC 1399&2T&$26.75\pm0.30$&$8.68\pm0.52$&$9.84\pm0.18$&$4.40\pm0.74$&
$3.83\pm0.30\pm0.92$&3.95\\
NGC 1550&2T&$33.85\pm0.49$&$11.61\pm0.93$&$10.63\pm0.27$&$9.38\pm1.28$&
$2.24\pm0.49\pm1.61$&1.33\\
NGC 2563&1T&$17.13\pm0.43$&$7.85\pm0.79$&$7.70\pm0.22$&$0.75\pm0.06$&
$0.82\pm0.43\pm0.82$&0.89\\
NGC 4325&1T&$13.07\pm0.49$&$6.62\pm0.79$&$5.69\pm0.21$&$0.71\pm0.03$&
$0.04\pm0.49\pm0.82$&0.04\\
NGC 5044&2T&$22.40\pm0.33$&$8.10\pm0.65$&$9.01\pm0.19$&$3.33\pm0.11$&
$1.96\pm0.33\pm0.68$& (2.58)\\
NGC 5846&2T&$25.15\pm0.59$&$12.59\pm1.01$&$10.47\pm0.34$&$0.35\pm0.03$&
$1.74\pm0.59\pm1.06$&1.43\\
NGC 6329&1T&$17.47\pm0.49$&$8.09\pm0.81$&$7.02\pm0.23$&$1.14\pm0.17$&
$1.22\pm0.49\pm0.84$&1.25\\
NGC 6521&1T&$13.57\pm0.44$&$5.90\pm0.71$&$5.64\pm0.19$&$1.52\pm0.44$&
$0.52\pm0.44\pm0.73$&0.60\\
NGC 7619&1T&$15.31\pm0.37$&$7.46\pm0.75$&$6.71\pm0.18$&$0.32\pm0.03$&
$0.81\pm0.37\pm0.77$&0.95\\
Pavo&2T&$10.20\pm0.42$&$5.60\pm0.67$&$4.11\pm0.15$&$0.51\pm0.25$&
$-0.02\pm0.42\pm0.73$&$-0.03$\\
RGH 80&1T&$10.94\pm0.29$&$4.61\pm0.55$&$4.44\pm0.11$&$0.34\pm0.05$&
$1.55\pm0.29\pm0.56$&2.45\\
S49-147&2T&$23.96\pm0.61$&$11.26\pm0.90$&$10.53\pm0.36$&$0.57\pm0.19$&
$1.59\pm0.61\pm0.99$&1.37\\
\end{longtable}

\begin{figure}[htbp]
	\begin{center}
	\FigureFile(80mm,50mm){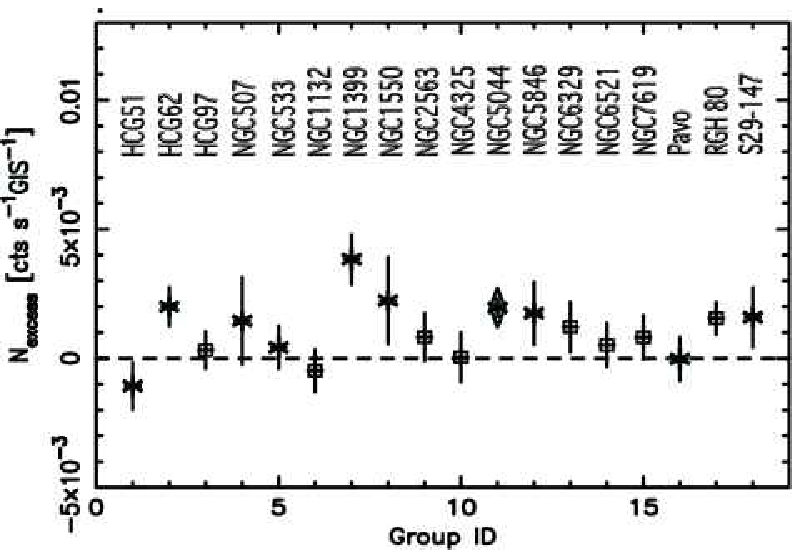}
	\end{center}
\caption{Significance of the excess count rate in the GIS 4--8 keV spectra,
above the IGM model determined in the energy range below 2.5 keV.
Open boxes show those objects of which the soft-band data 
are reproduced with the vMEKAL model,
while stars those which require  2-vMEKAL modeling.
Error bars refer to the quadrature sum of statistical and systematic 1-sigma uncertainties.
Result of NGC 5044, in which the soft-band fit is not acceptable in 99\% confidence, is shown with a diamond. 
}
\label{fig:other:anaspec:excess:sig}
\end{figure}

The 7th column in  table~\ref{tbl:other:excess:excess}
lists the 4--8 keV excess count rates of the 18 objects,
derived by subtracting  the CXB, NXB, and IGM from the raw data.
There, the statistical and systematic errors are given separately;
the former comes from that of the raw data,
while the latter is the quadrature sum of those 
associated with the three  components subtracted. 
The last column represent significance of the excess counts,
calculated against the overall uncertainty
which is the quadrature sum of the statistical and systematic errors.
These values are also presented in figure \ref{fig:other:anaspec:excess:sig}.
Thus, the hard-band excess above the soft-band determined IGM model
is insignificant ($\lesssim 1 \sigma$ level) in 10 out of the 18 objects,
and is marginal ($\sim 1.5 \sigma$ level) in 4 objects
(NGC 1550, NGC 5846, NGC 6329 and S49-147).
In contrast, the significance is higher than $2\sigma$
in 3 groups of galaxies, namely HCG 62, NGC 1399, and RGH 80.
Among them, the soft band IGM modeling of the first and the last 
objects is statistically acceptable.
In the case of NGC 1399, the somewhat poor IGM modeling
(acceptable at 98\%) hampers us to draw a firm conclusion,
but the hard excess could also be present because its nominal
significance is rather high ($4\sigma$).
Properties of these 3 objects are discussed in detail in the next sub-section.

The NGC 5044 group also shows excess significance as $2.6\sigma$.
The IGM modeling in this source, however, is unacceptable 
even by 99\% confidence (with $\chi^2$/dof = 116.7/79), 
so that the errors
associated with the IGM contribution could be underestimated.
In addition, the excess hard emission itself is weak.
We estimated the 2--10 keV hard band flux $F_{\rm hard}$ assuming 
a power-law model with photon index $\Gamma$ fixed at 2.0,
from the excess hard counts in table~\ref{tbl:other:excess:excess}.
The value of $F_{\rm hard}$ thus obtained amounts to only 4\% of the $F_{\rm soft}$.
Thus, we conclude that the hard excess signal
suggested in NGC 5044 is not significant enough in this study.

\subsection{Detailed analysis of selected objects}
\label{chap:spec:detail}

In this section, we analyze in detail the 3 objects,
HCG 62, NGC 1399 and RGH 80,
which have been found in section~4.3.2 to require a significant hard X-ray component, 
in addition to the thermal IGM  model (either vMEKAL or 2-vMEKAL).
As already reported by Fukazawa et al.~(2001), 
the compact group HCG 62 shows the strong excess signal, 
and the IGM contribution to its 4--8 keV excess counts is only $\sim$ 25 \%.
Therefore, this object provides an ideal benchmark test for our work,
particularly in comparison with Fukazawa et al.~(2001).
We study RGH 80 as well,
because it also shows evidence of strong hard excess signal, 
including the high value of $kT_{\rm H} = 2.62$ keV.
In the case of NGC 1399, the soft band IGM modeling is only marginally acceptable 
and also the value of $kT_{\rm H}$ is less than 2 keV,
while the excess is still significant in the previous analysis.
Thus, this objects also requires detailed analysis.

\begin{longtable}{lllllllll} 
\caption{Results from fitting to the full band spectra with a model with and without a power-law.}
\label{tbl:other:spec:fhard}
\hline \hline
target  & Model\footnotemark[$*$]  & $kT_1$\footnotemark[$\dagger$] &$kT_2$\footnotemark[$\ddagger$]  & Abundance   
 & $F_{\rm hard}$\footnotemark[$\S$]  &$F_{\rm soft}^{\rm IGM}$\footnotemark[$\|$] & $\chi^2$/dof  
 & $\Delta\chi^2$ \footnotemark[$\#$]   \\
            & & (keV)    & (keV)  & $Z_{\rm Fe}$($Z_{\odot}$) & &  & & \\
\hline
\endhead
\hline
\endfoot
\hline
\multicolumn{9}{@{}l@{}}{\hbox to 0pt{\parbox{160mm}{\footnotesize
     \par\noindent
     \footnotemark[$*$] Model ID. ``1T'' means vMEKAL, and ``2T'' means 2-vMEKAL models. 
     ``2Z'' means that the metal abudances are grouped into two ($Z_{\alpha}$ and $Z_{\rm Fe}$). 
     ``6Z'' means models with Mg, Al, Si and S set free, in addition to "2Z". 
     ``PL'' mean the $\Gamma = 2.0$ fixed power-law component.
          \par\noindent
     \footnotemark[$\dagger$] Temperature of the cooler component. 
          \par\noindent
     \footnotemark[$\ddagger$] Temperature of the hotter component, when fitted with the 2-vMEKAL model.
          \par\noindent
     \footnotemark[$\S$] 2-10 keV flux of the power-law component, in $10^{-12}$ erg s$^{-1}$ cm$^{-2}$.
     The second error represents the value of 90\% CXB fluctuation.
    \par\noindent
     \footnotemark[$\|$] 0.7-2.5 keV flux of the IGM component. ($10^{-12}$ erg s$^{-1}$ cm$^{-2}$)
     \par\noindent
    \footnotemark[$\#$] Improvement of $\chi^2$ by adding a power-law component (addition of 2 parameters).
      \par\noindent
          }\hss}}
\endlastfoot
HCG 62&2T/2Z&$0.94_{-0.04}^{+0.04}$&$17.5_{-11.8}^{+\infty}$&$0.17_{-0.01}^{+0.04}$&
- &4.7&84.4/66 & - \\
&2T/2Z+PL&$0.71_{-0.08}^{+0.12}$&$1.20_{-0.14}^{+0.33}$&$0.30_{-0.07}^{+0.15}$&
$1.34_{-0.26}^{+0.25}\pm 0.53$&4.4&61.3/64 & 23.1\\
NGC 1399&2T/6Z&$0.88_{-0.03}^{+0.19}$&$1.79_{-0.09}^{+0.21}$&$0.78_{-0.15}^{+0.08}$&
- &17.3&224.8/173 & -\\
&2T/6Z+PL&$0.84_{-0.06}^{+0.04}$&$1.58_{-0.13}^{+0.11}$&$0.63_{-0.11}^{+0.12}$&
$1.90_{-0.61}^{+0.64} \pm 0.45$&16.5&195.2/171 & 29.6\\
RGH 80&1T/2Z&$1.26_{-0.05}^{+0.05}$& - &$0.22_{-0.04}^{+0.06}$&
- &1.8&111.5/66 &- \\
&1T/2Z+PL&$1.08_{-0.04}^{+0.03}$& - &$0.32_{-0.10}^{+0.21}$&
$0.69_{-0.14}^{+0.13} \pm 0.37$&1.3&58.2/64 &53.3 \\
\end{longtable}

\subsubsection{The HCG 62 group}
\label{chap:spec:detail:h62}

This is a bright group of galaxies,
emitting a 0.7--2.5 keV flux of $4.7\times 10^{-12}$ erg s$^{-1}$ cm$^{-2}$.
To investigate the nature of the excess hard emission in this source,
we analyzed the full band spectra.
The GIS spectra were fitted over the 1--8 keV range, 
while those of the SIS were used up to 4.5 keV 
determined from the source brightness and background uncertainty.

\begin{figure*}[htbp]
	\begin{center}
	\FigureFile(160mm,70mm){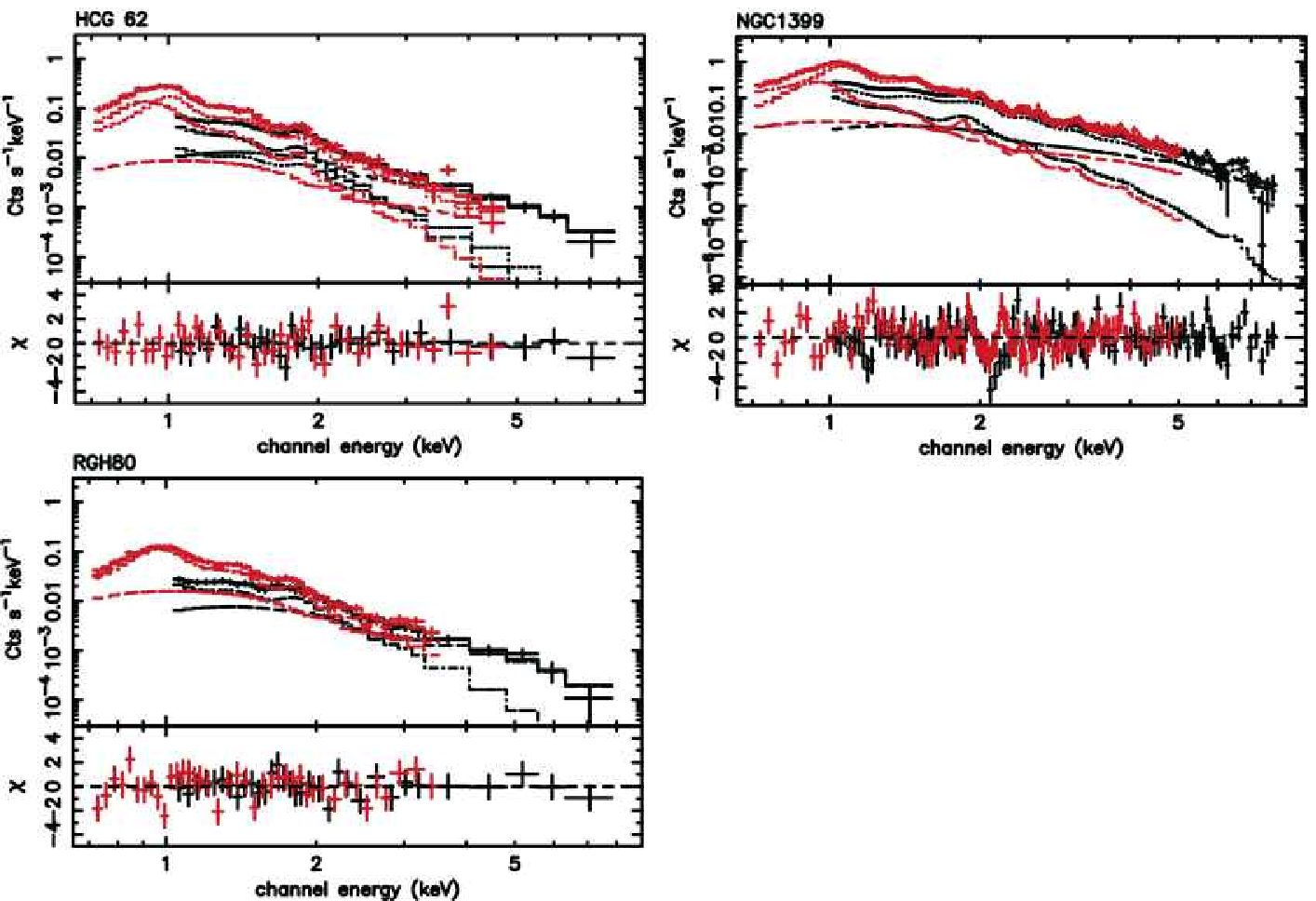}
	\end{center}
\caption{Results of the joint 2-vMEKAL+PL model fitting to the GIS and SIS spectra 
of HCG 62 and NGC 1399. Those of RGH 80 fitted with vMEKAL+PL model is also presented. 
See text for detail.}
\label{fig:h62:sp:six_spec}
\end{figure*}

The 2-vMEKAL model provided unacceptable fit to the full band data
with $\chi^2$/dof = 84.4/66.
By adding a power-law component with a photon index
$\Gamma$ fixed at 2.0 (2-vMEKAL+PL model),
an acceptable fit was obtained with $\chi^2$/dof = 61.3/64.
Accorring to an $F$-test,
the probability of this improvement being by chance is less than $10^{-4}$.
The results of both fits are presented in 
table~\ref{tbl:other:spec:fhard}, and the fit with the latter model is 
presented in figure~\ref{fig:h62:sp:six_spec}.
The power-law flux in the 2--10 keV band, $F_{\rm hard}$, 
amounts to  $\sim30$\% of
the IGM component flux in the 0.7--2.5 keV band, $F_{\rm soft}^{\rm IGM}$. 
These results generally agree with Fukazawa et al.~(2001).
The excess above the 2-vMEKAL fit  can also be reproduced
by a bremsstrahlung model instead of the power-law,
although the  bremsstrahlung temperature becomes
rather high as $4.0_{-1.3}^{+8.3}$ keV.
Thus, the data clearly requires a non-thermal, 
or less likely, very hot thermal component.
If  $\Gamma$ is  set free in the above 2-vMEKAL+PL fit, 
we obtain $\Gamma = 2.17_{-0.53}^{+0.28}$ with $\chi^2/$dof = 60.8/63,
implying that the improvement is insignificant.
For reference, the 90\% fluctuation level of the CXB brightness is 
separately presented in the error column of $F_{\rm hard}$ in table ~\ref{tbl:other:spec:fhard}.
Again, the CXB fluctuation is insufficient to explain the hard excess,
confirming the results presented in section 4.3.

\begin{longtable}{lllllllll} 
\caption{Fit results to the radially sorted SIS+GIS spectra of HCG 62, with the model including power-law.}
\label{tbl:h62:sp:gsr0-60vmekals}
\label{tbl:h62:sp:gsrradimekalpo20}
\label{tbl:h62:sp:gsrcenvmekalmt2po20}
\hline \hline
Region & $kT_1$ \footnotemark[$*$]  &$ kT_2$ \footnotemark[$\dagger$]  &  \multicolumn{2}{l}{Abundance}  
&  $F_{\rm hard}$ \footnotemark[$\ddagger$]  & $ F_{\rm soft}^{\rm IGM}$ \footnotemark[$\S$] 
& $\chi^2$/dof  &$\Delta\chi^2$ \footnotemark[$\|$]     \\
              & (keV)       & (keV)   & $Z_{\alpha}$&$Z_{\rm Fe}$  & 
               \multicolumn{2}{l}{ } & & \\
\hline
\endhead
\hline
\endfoot
\hline
\multicolumn{9}{@{}l@{}}{\hbox to 0pt{\parbox{160mm}{\footnotesize
     \par\noindent
     \footnotemark[$*$] Temperature of the IGM model. In 2-vMEKAL models, a cooler component value is presented.
     \par\noindent
     \footnotemark[$\dagger$] Temperature of the hotter component in 2-vMEKAL models.
     \par\noindent
     \footnotemark[$\ddagger$] The 2-10 keV flux of the power-law emission. ($10^{-12}$ erg s$^{-1}$ cm$^{-2}$)
     \par\noindent
     \footnotemark[$\S$] The 0.7-2.5 keV flux of the IGM emission.($10^{-12}$ erg s$^{-1}$ cm$^{-2}$)
     \par\noindent
     \footnotemark[$\|$] Improvement of $\chi^2$ by adding a power-law component (addition of 2 parameters).
     In all cases, the improvement were significant by ($1/1000$) by F-test.
      \par\noindent
 }\hss}}
\endlastfoot
$0'<r<3'$       & $0.87_{-0.01}^{+0.03}$& - & $0.69_{-0.24}^{+0.25}$&$0.34_{-0.10}^{+0.19}$ 
& $0.26_{-0.04}^{+0.04}$ & 1.11 &77.3/66 & 79.3\\
& $0.77_{-0.14}^{+0.05}$&$1.42_{-0.34}^{+0.27}$&$1.26_{-0.63}^{+2.43}$&$0.69_{-0.27}^{+1.11}$ 
& $0.25_{-0.04}^{+0.07}$ &1.24&55.8/63& 23.9\\
$3'<r<7'.5$     & $1.10_{-0.05}^{+0.18}$& - & $0.26_{-0.11}^{+0.13}$&$0.16_{-0.08}^{+0.14}$
 & $0.30_{-0.08}^{+0.08}$ & 1.06&58.6/64 & 33.2\\
$7'.5<r<15'$  & $0.84_{-0.09}^{+0.16}$& - & $0.13_{-0.13}^{+0.26}$&$0.15_{-0.06}^{+0.11}$
 & $0.66_{-0.16}^{+0.14}$ & 1.72& 59.8/58 & 37.6\\
\hline
$3'<r<15'$     & $0.97_{-0.09}^{+0.06}$& - & $0.19_{-0.11}^{+0.15}$&$0.17_{-0.04}^{+0.06}$
& $0.92_{-0.17}^{+0.18}$ & 2.71  &73.2/65& 67.8\\
\end{longtable}

\begin{figure}[htbp]
	\begin{center}
	\FigureFile(80mm,50mm){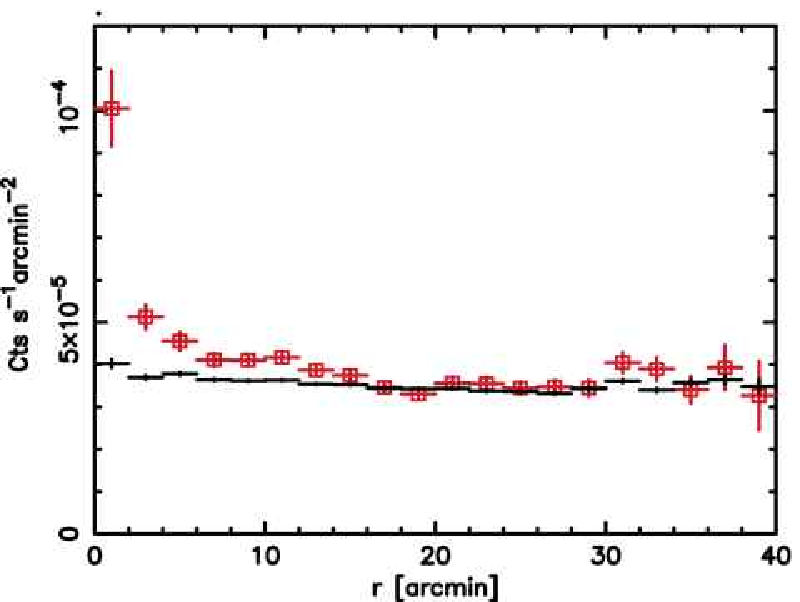}
	\end{center}
\caption{The 4.0--8.0 keV  GIS radial count-rate profile,
shown as a function of two-dimensional radius  from the group center.
The on-source data are shown in red,
and the background in black.
Abscissa is in arcmin, while ordinate is in cts s$^{-1}$ arcmin$^{-2}$.}
\label{fig:h62:signi:pror500-900}
\end{figure}

Figure~\ref{fig:h62:signi:pror500-900} shows the azimuthally-averaged 
radial profile of the 4.0--8.0 keV GIS count rate from HCG 62,
compared with the background profile obtained as described in section \ref{chap:screen}.
Regions around the five point sources in the HCG 62 field are excluded.
Thus, the comparison reconfirms the highly significant hard X-ray signal,
of which only $\sim 25$\% can be accounted for by the thermal IGM emission
as revealed by figure~\ref{fig:h62:sp:six_spec}.
The emission is clearly extended and detectable up to $\sim 15'$,
beyond which it vanishes in agreement with the background 
scaling correction employed in section \ref{chap:screen}.

To further examine the consistency between 
the spectral (figure~\ref{fig:h62:sp:six_spec}) 
and spatial (figure~\ref{fig:h62:signi:pror500-900}) results, 
we sorted the spectra into three concentric annuli,
and performed the 2-vMEKAL+PL fit to each.
As described in table \ref{tbl:h62:sp:gsrradimekalpo20},
the power-law component is required in all annuli.
The centeral $3'$ of the group is well fitted with a vMEKAL+PL model,
but 2-vMEKAL+PL fit provides significantly 
improved results in view of an $F$-test.
The outer regions can be fitted with a single vMEKAL model
on condition that the $\Gamma=2.0$ power-law is added.
Therefore, the extended nature of the excess hard X-ray emission,
indicated by figure~\ref{fig:h62:signi:pror500-900},
is supported  by the spectral analysis as well.

Buote (2000)  analyzed the same ASCA data 
accumulated from the central $\sim 3'$ region,
and fitted the spectra with a combination of 
0.7 keV and 1.4 keV thermal components 
with $\sim 1$ solar metal abundances.
While the reported temperatures agree with our measurements,
he did not mention  the excess hard component in his paper,
possibly because he used the energy range below $\sim 5$ keV.
This limited energy range, in turn, was required presumably
by a more conventional background subtraction method employed.
Actually, we can reproduce his results
if we ignore the energy range above 4 keV 
and  set the absorption column free, as he did.

Recently, Morita et al.~(2006) analyzed the XMM-Newton and Chandra data of HCG 62,
and obtained a radial temperature profile 
which is $0.7-0.8$ keV within $1'$  
and is increasing to 1.4 keV at $2'-4'$. 
Again, the measured temperatures  are consistent 
with our 2-vMEKAL+PL results from the $r < 3'$ spectra.
They did not detect the excess hard X-rays in their data. 
Instead, they included the emission by scaling the parameters of Fukazawa et al. (2000)
as a background in their analysis.
They argue that their results did not significantly change with and without this component.
Non detection by Chandra and XMM is not surprising,
since their 4--8 keV background count rate,
normalized to the effective area and the sky area,
is  8 and 4 times higher than that of the GIS, respectively
(figure \ref{fig:intro:GISmerit}).
Specifically, the hard excess emission from HCG 62 amounts to 
12\% of the 4--8 keV GIS background (CXB+NXB),
whereas it is only $\sim 3$\% of the total background of XMM;
this is smaller than the background modeling uncertainty of $\sim 5$\%,
obtained after sophisticated screening (e.g. Nevalainen et al. 2005).

\subsubsection{The NGC 1399 group}
\label{chap:spec:ngc1399}

Since the NGC 1399 group (the Fornax cluster) is 
one of the X-ray brightest groups in the sky,
the data have very high statistics, requiring careful analyses.
The soft-band spectra can be approximated by two temperatures 
of 0.8 keV and 1.4 keV  (table \ref{tbl:other:excess:2vmekal}),
which are consistent with the virial temperature of $\sim 1$ keV
implied by the galaxy velocity dispersion of 374 km s$^{-1}$ (Drinkwater et al. 2001).
Compared to these, the hard-band derived temperature,
$kT_{\rm H} = 1.84_{-0.11}^{+0.09}$ keV (table \ref{tbl:other:spec:single}),
is significantly higher, suggesting the presence of a harder emission component.

As an attempt to further examine the issue, we started with fitting
the 2-vMEKAL model to the full band spectra.
The abundances of Mg, Al, Si, and S were set free in addition 
to $Z_{\alpha}$ and $Z_{\rm Fe}$.
The fit was far from acceptable (with $\chi^2/$dof = 292.1/191), 
due to the possible calibration discrepancies
around ~1.2 and ~2.1 keV  mentioned in section 4.2.4.
By ignoring the energy bands of 1.15--1.25 keV and 2.1--2.2 keV,
the 2-vMEKAL model fit improved to $\chi^2/$dof=224.8/173,
but was still unacceptable by 99.5\%.
By further adding a power-law with fixed $\Gamma=2.0$  (2-vMEKAL+PL model),
we have finally obtained an acceptable fit by 90\% level (with $\chi^2/$dof=195.2/171).
The results are
presented in table~\ref{tbl:other:spec:fhard} and figure~\ref{fig:h62:sp:six_spec}.
(In the figure, the two ignored energy bands are restored for clarity.)
The power-law flux $F_{\rm hard}$ is 12\% of $F_{\rm soft}^{\rm IGM}$.

In the 2-vMEKAL fit after ignoring the two energy bands,
the best fit value of the hotter component temperature 
is derived as $kT_2 = 1.79$ keV.
Since it is close to $kT_{\rm H} = 1.84$, 
and since the 2-vMEKAL fit is almost successful,
a careful examination is needed
to judge if the data is really requiring the hard component.
The improvement by adding a power-law component is 
$\Delta \chi^2 = -29.6$, 
which is significant with the chance probability of less than $10^{-5}$ in terms of an $F$-test.
Furthermore, the resultant fit with 2-vMEKAL+PL model is statistically acceptable.
Therefore, the excess hard X-ray emission is likely to be present
in the NGC 1399 group as well.
If we replace the power-law with a third thermal component,
a similarly good ($\chi^2/$dof=189.6/170) full-band fit can be obtained.
In this case, however, the temperature become as high as $3.2_{-0.9}^{+2.6}$ keV. 
Thus, the excess hard component in this object 
can be explained by a $\Gamma=2.0$ power-law emission
which is possibly of non-thermal nature,
or a very hot  thermal emission  with a temperature of  $\gtrsim 3$ keV.

\subsubsection{The RGH 80 group}
\label{chap:spec:rgh80}

The last of the 3 selected objects, RGH 80 is the most distant one in our sample,
yet showing rather strong hard X-ray signals above the extrapolated IGM contribution.
The full band fit with a single vMEKAL model was not acceptable 
(with $\chi^2$/dof = 111.5/66).
When we added a power-law component with $\Gamma$ again fixed at 2.0,
the fit greatly improved and became acceptable with $\chi^2$/dof = 58.2/64
(see table~\ref{tbl:other:spec:fhard} and figure~\ref{fig:h62:sp:six_spec}). 
Although the absolute value of $F_{\rm hard}$ itself is rather low, it amounts to
$\sim 50\%$ of $F_{\rm soft}^{\rm IGM}$.
When we added a second thermal component in place of the power-law (2-vMEKAL model),
a similarly good fit was obtained with $\chi^2$/dof = 58.7/63.
However, the obtained hotter temperature  is
$kT_2 = 11.0_{-8.3}^{+\infty}$ keV, that is, higher than 2.7 keV.
Since this value is considerably higher than the cooler one
($kT_1 = 1.09_{-0.04}^{+0.07}$ keV)
and much exceeding typical virial temperatures of galaxy groups,
the excess hard signals are likely to be of non-thermal origin. 
These properties make this object resemble HCG 62,
but with poorer statistics.

Boute (2000) analyzed the ASCA spectra of RGH 80
derived from the central $\sim 3'.6$,
using also a two temperature thermal emission model,
to obtain a hotter temperature of $kT_2 = 1.64_{-0.17}^{+0.21}$ keV.
Although this is apparently inconsistent with our results,
we confirmed that we can reproduce the Buote's result
by extracting the spectra from a similar region.
Therefore, the hard emission is inferred to be stronger in the $3'<r<10'$ region.

\section{Discussion}
\label{chap:discussion}

\subsection{Summary of analysis results}
\label{chap:discussion:overall}

Through ASCA observations of 18 near-by low temperature galaxy groups,
indication of excess hard component was obtained from 3 objects.
The excess first manifested itself as
large differences between the temperatures inferred
in the soft band below 2.5 keV ($kT_{\rm S}$) and in the hard band above 2.5 keV ($kT_{\rm H}$), 
which are determined mainly by the Fe-L emission lines
and  the bremsstrahlung continuum, respectively.

In order to quantify the suggested hard-band excess,
we represented the IGM emission by 
vMEKAL or 2-vMEKAL model determined in the energy range below 2.5 keV,
and extrapolated it to the 4--8 keV GIS range.
Then, even taking fully into account the CXB fluctuation and the NXB estimation error, 
three objects (HCG 62, NGC 1399, and RGH 80)
showed significant ($> 2\sigma$) excess counts above the 
expected IGM contribution (section~\ref{chap:spec:excess:n_excess}).
From the full band fitting to these objects (section~\ref{chap:spec:detail}), 
we found that the excess can be successfully represented
by a $\Gamma =2.0$ power-law model, and hence it is likely to be
of non-thermal origin, particularly in HCG 62 and RGH 80.
The spectra of NGC 1399 can be explained either by 
adding a non-thermal emission with a flux of about 10\% of that of the IGM,
or the third thermal component having 
a temperature $\gtrsim 3$ keV.

\subsection{Hard X-ray excess compared to other parameters}
\label{chap:discussion:compare}

\begin{longtable}{llll} 
\caption{Luminosity of the IGM and excess emission of HCG 62, NGC 1399 and RGH 80.
Estimated contribution from LMXBs in the member galaxy is also shown.}
\label{tbl:discussion:overall:1}
\hline \hline
target & $L_{\rm soft}^{\rm IGM}$ \footnotemark[$*$] & $L_{\rm hard}$ \footnotemark[$\dagger$] & $L_{\rm LMXB}$ \footnotemark[$\ddagger$] \\
\hline
\endhead
\hline
\endfoot
\hline
\multicolumn{4}{@{}l@{}}{\hbox to 0pt{\parbox{80mm}{\footnotesize
     \par\noindent
     \footnotemark[$*$]  Luminosity of the IGM emission in 0.7-2.5 keV, in unit of $10^{40}$ erg s$^{-1}$. 
          \par\noindent
     \footnotemark[$\dagger$] Luminosity of the excess hard emission in 2-10 keV, in unit of $10^{40}$ erg s$^{-1}$. Errors are $1\sigma$ including both the statistical and systematic origins.
         \par\noindent
     \footnotemark[$\ddagger$] Luminosity of the expected LMXB emission in 0.7-2.5 keV, in unit of $10^{40}$ erg s$^{-1}$.
     \par\noindent
 }\hss}}
\endlastfoot
HCG 62&181.2&$55.3\pm14.6$&2.5\\
NGC 1399&73.7&$8.5\pm2.1$&1.8\\
RGH 80&341.1&$182.4\pm63.7$&1.3\\
\end{longtable}

\begin{figure}[htbp]
	\begin{center}
	\FigureFile(80mm,60mm){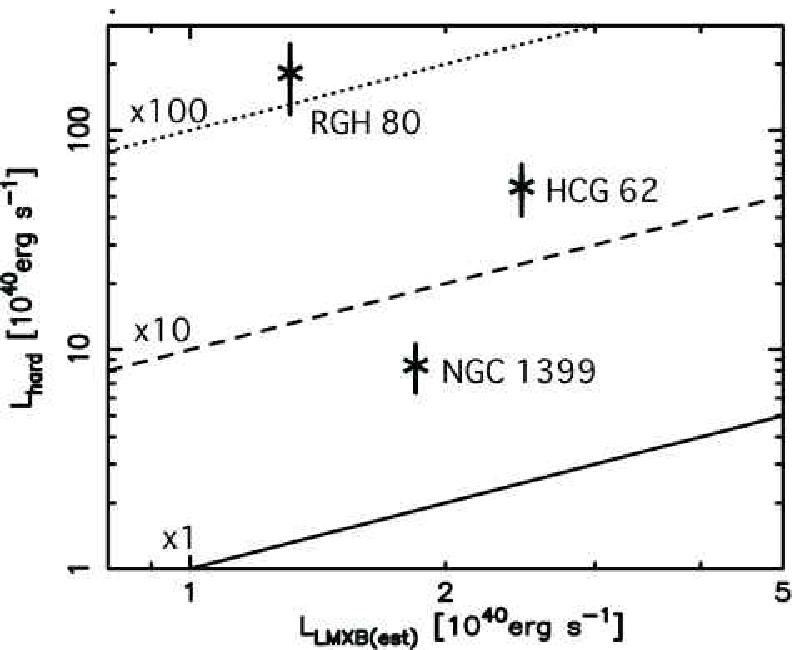}
	\end{center}
\caption{Excess hard luminosity, $L_{\rm hard}$, compared with $L_{\rm LMXB}$
which is derived from the X-ray to optical flux ratios of elliptical galaxies.
Solid, dashed and dotted lines represent, 1, 10 and 100 times of  $L_{\rm LMXB}$, respectively.
}
\label{fig:other:anaspec:excess:lxlb}
\end{figure}

In section~\ref{chap:spec:detail}, we derived the value of $F_{\rm hard}$
and the associated errors for HCG 62, NGC 1399 and RGH 80.
In this section, we briefly compare it to other parameters of these objects.
The calculated 2--10 keV luminosity of the power-law component,
$L_{\rm hard}$, is presented in table \ref{tbl:discussion:overall:1},
together with the 0.7-2.5 keV luminosity of the IGM component ($L_{\rm soft}^{\rm IGM}$).
Both the values of $F_{\rm soft}$ and $kT_{\rm S}$ of 
the three objects are typical in our sample.
In addition, some other objects in our sample have similar temperatures and
IGM luminosities to the three object, yet without excess hard signals.
The lack of correlations to these parameters suggests that the phenomenon
we have been studying is not likely to be artifacts caused by
wrong background subtraction, or incorrect modeling of the IGM contribution.

An elliptical galaxy is known to emit an X-ray component with a rather hard spectrum,
as a sum of discrete sources in it, such as LMXBs in particular
(Canizares et al.~1987; Matsushita et al.~1994; Matsushita et al.~2000).
This component could provide a possible explanation to the hard excess emission,
because LMXB spectra,
approximated by a thermal bremsstrahlung with $kT \sim 10$ keV,
are generally consistent with what has been observed in the present study.
The  integrated 2--10 keV discrete-source luminosity, $L_{\rm LMXB}$,
of each elliptical galaxy is known to be approximately
proportional to its  B-band luminosity, $L_B$. 
We hence estimated $L_{\rm LMXB}$ in our sample objects,
using $L_B$ given in table 1 and the relation of 
$L_{\rm LMXB} = 4\times10^{39} (L_B/10^{10}L_{\odot})$ erg s$^{-1}$
(converted from Matsushita et al.~2000).
The result, presented in figure \ref{fig:other:anaspec:excess:lxlb},
shows that $L_{\rm hard}$ of HCG 62, NGC 1399 and RGH 80
is 20, 5 and 140 times higher than the estimated $L_{\rm LMXB}$, respectively.
Thus, the discrete-source contribution cannot explain away the hard-excess phenomenon.
Similar excess hard X-ray fluxes from a few Virgo elliptical
galaxies were reported by Loewenstein et al. (2001).

\subsection{Possible Emission Mechanisms}
\label{chap:discuss:emission}

Even in the case of HCG 62 
which shows the most significant  hard X-ray excess,
the spectral shape of the excess emission is not well constrained;
$\Gamma = 2.17_{-0.53}^{+0.28}$.
Therefore, it is rather difficult to tell whether the emission is of thermal or non-thermal origin.
In this section, we briefly discuss  both scenarios,
using HCG 62 as a representative case.

\subsubsection{Non-thermal interpretation}
\label{chap:discuss:emission:nth}

In the non-thermal scenario, two possibilities are generally discussed:
inverse Compton (IC) emission from GeV electrons
as they scatter off the cosmic microwave background photons,
and non-thermal bremsstrahlung from sub-relativistic particles
interacting with ambient plasmas (see e.g. Sarazin 1999 and Sarazin, Kempner 2000).
However, the latter is unlikely, because of too low an efficiency (e.g. Petrosian 2001);
sub-relativistic electrons suffer from 4 to 5 orders of magnitude larger energy loss 
in their Coulomb interactions with ambient ions, than their bremsstrahlung loss,
making the energetics unrealistic (e.g. Fukazawa et al.~2001).
Therefore, below we consider only the IC interpretation.

In the IC scenario, the postulated GeV electrons should also
emit synchrotron photons in the radio band.
There is however no reported radio halo detection of HCG 62, 
and the 365 MHz Texas catalog (Douglas et al.~1996) gives 
an upper limit of $\sim 0.4$ Jy on the radio flux density from HCG 62.
The comparison of this upper limit with our 
$F_{\rm hard} \sim 1.3\times 10^{-12}$ erg s$^{-1}$ cm$^{-2}$ in the 2-10 keV band
yields an upper limit on the volume-averaged
magnetic field as $B\sim 0.1~\mu$G,
assuming that a single population of electrons with an energy index of 3.0
are emitting both IC and synchrotron components under a uniform magnetic field.
As already mentioned in Fukazawa et al.~(2001),
this limit appears too low for intra-group magnetic fields.
Introduction of non-uniform magnetic fields and/or time evolution of 
the electron energy distribution (e.g. high-energy cutoff) may solve this discrepancy.
For example, Brunetti et al. (2001) proposed 
a model explaining the non-thermal signature of the Coma cluster.
In this cluster, observed radio halo flux and the hard X-ray flux suggested by Beppo-SAX 
(e.g. Fuso-Femiano et al. 1999) leeds to a similarly low magnetic 
field ($\sim 0.1~\mu$G) when a simple model for electron population is employed.
By incorporating the radial dependence of
magnetic field and assumed (re)-acceleration power,
as well as the time evolution, in particular introducing the re-acceleration phase
which modifies the electron spectra flatter with distinctive cut-off, 
they successfully reproduced the Coma results.
Similar models may be able to explain the HCG 62 results.

When  $B< 3~\mu$G, the electrons are expected to lose their energies predominantly in the IC channel.
Therefore, to sustain $L_{\rm hard} \sim 4\times 10^{41}$ erg s$^{-1}$
under a steady-state condition,
a comparable energy input  should be supplied to the  electrons.
The recent Chandra detection of a pair of ``X-ray cavities" 
near the central galaxy (NGC 4761) of HCG 62
(Vrtilek 2000 \footnote{http://chandra.harvard.edu/photo/2001/hcg62/}, Morita et al.~2006)
suggests a past AGN activity, which may well have supplied the needed energy input.
Although NGC~4761 currently shows little evidence of AGN activity,
the scenario remains intact if the putative AGN activity continued
till 1 Gyr ago or later, because the cooling time of a 1 GeV electron
due to the IC process is $\sim 1$ Gyr.

Since HCG 62 is a compact group with a high galaxy density
and a rather large velocity dispersion of $376_{-46}^{+52}$ km s$^{-1}$ (Zabludoff, Mulchaey 1998),
energy inputs may also be possible from magneto-hydrodynamic interactions
of the member galaxies with the IGM (Makishima et al.~2001).
Thus, the non-thermal interpretation is promising
from the energetics view point.

\subsubsection{Thermal interpretation}
\label{chap:discuss:emission:th}

If the hard excess  in our sample objects are interpreted as thermal emission,
the inferred temperature ranges from $\sim$ 3 keV or higher.
In order to explain $L_{\rm hard}$ that amounts to up to 25\% of $L_{\rm soft}$,
then the putative hotter gas must fill $>70$\% of the group volume,
assuming that it is in a pressure balance with the $\sim$ 1 keV IGM.
That is,  the hotter gas is implied to be energetically dominant
in the intra-group space. Since the velocity dispersion of HCG 62 as quoted above
translates to a virial temperature of only $\sim 1$ keV (e.g. Xue, Wu, 2000),
the  postulated gas is concluded to be significantly hotter 
than the gravitational potential felt by the galaxies.
However, if the gas were gravitationally unbound and freely escaping with sound velocity, 
the necessary energy input would become enormous,
because the escape time of a 2 keV gas is two orders of magnitude
shorter than its radiative cooling time,
assuming a representative density of $\sim 3 \times 10^{-4}$ cm$^{-3}$.

Another possibility is that such an object is surrounded by a much deeper gravitational potential halo
with a considerably larger scale.
Actually, the presence of such a large-scale halo has been suggested
by Matsushita et al.~(1998) around the elliptical galaxy NGC~4636.
In any case, the presence of such a hot gaseous component
would have a profound impact on the structure and formation of galaxy groups.

\section{Conclusion and future prospect}
\label{chap:conclusion}

From the detailed analysis of the hard ($> 2.5$ keV) band
X-ray spectra of groups of galaxies obtained with ASCA,
evidence of excess hard X-ray emission
has been suggested from 3 out of 18 objects investigated.
They are HCG 62, NGC 1399 and RGH 80.
The emission cannot be explained by
either fluctuations in the CXB brightness,
the NXB estimation error, or contributions of point sources in the member galaxies.
The excess cannot be explained away
by assuming a moderate temperature gradient in the IGM, either.
At least in HCG 62, the hard X-ray emission is
as extended as the thermal IGM emission.

The observed excess hard X-ray emission can be modeled
by a power law with photon index fixed at 2.0, or a high-temperature thermal component,
although it is difficult to distinguish them.
If considered to be of non-thermal origin,
the observed hard X-ray emission can be most
reasonably interpreted as inverse-Compton emission
by GeV electrons accelerated in these systems.
In contrast, thermal interpretation of the
phenomenon leads to an inference
that some of groups of galaxies are surrounded
by a much deeper (and probably of larger-scale)
gravitational halo than those felt by the member galaxies.

In order to further promote the study,
we may utilize the Suzaku mission,
launched into orbit on 10th July, 2005.
Actually, the X-ray Imaging Spectrometer onboard the satellite
has a 6 times larger effective area than the ASCA GIS at 7 keV in total,
and its background, when normalized to the solid angle and effective area,
is only slightly higher than that of the GIS.

\bigskip

The authors a greatly thankful to the anonymous referee for critical reading and providing
fruitful comments on this work.

\end{document}